\documentclass[article]{aa}
\usepackage{epsfig}
\usepackage{ulem}
\usepackage{graphics}
\usepackage{amssymb}
\input colordvi      
\begin{document}
          
%\thesaurus{ 8
%            (09.03.1; %ISM: clouds
%             09.13.2; %ISM: molecules
%             09.19.1; %ISM: structure
%             13.18.3; %Continnum ISM
%             13.19.3) %Line ISM
%  }

\title{Radio-millimetre investigation of galactic infrared dark clouds}

\author{D. Teyssier, P. Hennebelle \and M. P\'erault}               

\offprints{D.~Teyssier \\
{\it e-mail:} teyssier@lra.ens.fr}

\institute{ Laboratoire de radioastronomie millim{\'e}trique, URA 336 du
CNRS, {\'E}cole normale sup{\'e}rieure and Observatoire de Paris, 24 rue Lhomond, 75231 Paris cedex 05,
France
}
\authorrunning{D. Teyssier et al.}
\titlerunning{Radio-mm investigation of galactic infrared dark clouds}
%\titlerunning{Radio-Millimetre Investigation of Galactic Infrared Dark Clouds}
\date{Received 29 December 2000 / Accepted 13 November 2001}

\abstract{
We present follow-up observations of the mid-Infrared dark clouds selected from the ISOGAL inner Galaxy sample. On-the-fly maps of $^{13}$CO, C$^{18}$O and the 1.2 mm continuum emission were conducted at the IRAM 30-m telescope, showing spectacular correlation with the mid-IR absorption. The dark clouds are distributed as far as the prominent molecular ring at a distance of 3 to 7 kpc from the Sun. The clouds exhibit shapes ranging from globules to thin filaments down to $\lesssim$ 1 pc in size. The on-the-fly images obtained in $^{13}$CO and C$^{18}$O confirmed that the cores are dense, compact molecular emitters, significantly more massive than local dark clouds (more than 1000 $\rm M_\odot$) and lie within low activity Giant Molecular Clouds (GMC's). Ratios of the emission in the $J=(2-1)$ and $(1-0)$ transitions of $^{13}$CO and C$^{18}$O show a remarkable uniformity within each cloud, with a significant portion of the sample represented well by a ratio of 0.67$\pm0.12$. Preliminary analysis of temperature and density measurements reveals that most of the cores have densities above $10^{5}$ cm$^{-3}$ and temperatures between 8 and 25 K, these latter clouds being associated with young embedded stars. Despite the high extinction inferred from mid-IR ($A_{\rm v} > 50$, Hennebelle et al. 2001), the molecular lines are surprisingly weak, indicating likely depletion onto cold grains.
}

\maketitle

\section{Introduction}
Galactic dark clouds observed in the mid-Infrared were first surveyed by ISO
%infrared dark clouds were first observed by ISO 
(P\'erault et al. 1996), and turn out to be the highly condensed parts of Giant Molecular Clouds (GMC's) kiloparsecs away from the Sun.
The ISOGAL survey imaged $\sim$ 10\% of the inner Galactic ridge, mostly towards relatively quiescent areas, in broad filters around 7 and 15 $\mu$m (Omont et al. 1999). A systematic analysis of the ISOGAL plates (Hennebelle et al. 2001, hereafter Paper I) allowed extraction of a catalogue of about 450 objects, most of them located in the inner Galaxy. The features are associated with GMC's lying between 3 and 7 kpc from us. In Paper I, we derived opacities at 15 $\mu$m in the range 1 to 4 for a few selected objects, leading to column densities of the order of 10$^{23}$ cm$^{-2}$.

Similar findings have been reported by Egan et al. (1998) from the MSX (Midcourse Space Experiment) survey of the whole Galactic plane. They counted about 2000 IR dark clouds (IRDC's) located in the same distance range as the ISOGAL clouds and estimated extinctions at 8 $\mu$m in excess of 2. Detections of few clouds at millimetre wavelengths (Carey et al. 1998) confirmed that these objects are dense ($n_{\rm H_2} > 10^{5}$ cm$^{-3}$) and cold (T $<$ 20 K). The authors conclude that these clouds are pre-protostellar cores where no sign of star-formation has been detected so far. This analysis was recently refined using continuum detection at 850 and 450 $\mu$m (Carey et al. 2000). The relatively high masses inferred from the submillimetre observations suggest a significant potential for stellar formation in these cores.

In the present paper we analyse spectroscopic and continuum follow-up observations of the ISOGAL dark features, conducted at the IRAM 30-m telescope. A sample of 13 objects has been mapped in several molecular tracers. A few characteristic physical parameters are inferred. The spatial resolution of these observations (11\arcsec~at 1.3 mm) opens the way to the analysis of the gas associated with the dense dust revealed by the mid-IR absorption at an intermediate spatial scale. Our main purpose is to compare the properties of our objects to those of the well-known local dark clouds. A subsequent study will provide an extensive analysis of temperature and density measurements and better assess the physico-chemical processes at work in the clouds (Teyssier et al. 2001b).
 
Sect.~\ref{obs} presents the observational conditions and strategy.
%[\sout{ details some of the original mapping and data reduction strategies developed in the scope of this project.}]
In Sect.~\ref{obs properties} we describe the correlation between the millimetre emission and the IR absorption data. The nature of the different objects is assessed and a detailed spatial and spectral analysis is introduced. Sect.~\ref{properties} gives the physical properties of the clouds and their relation to the mid-IR opacities estimated in Paper I. Sect.~\ref{conclusion} summarises our conclusions.

%%%%%%%%%%%%%%%%%%%%%%%%%%%%%%%%%%%%%%%%%%%%%%%%%%%%%%%%%%%%%%%%%%%%%%%%%%%%%%%

\begin{table}
\begin{center}
\caption{Line parameters}
\label{line param}
\begin{tabular}{l r r r r} 
\hline \hline \\
Species & Transitions & $\nu_{\rm 0}$ & HPBW & $T_{\rm sys}$ \\
  &  & (GHz)   &(\arcsec)     & (K) \\  \hline
$^{13}{\rm CO}$   & $1 \rightarrow 0$ & 110.201353 &  22.5  & 150-250  \\
                  &  $2 \rightarrow 1$ & 220.398686 & 11.2  & 400-900  \\
${\rm C^{18}O}$   & $1 \rightarrow 0$ & 109.782160  & 22.5  & 150-250 \\ 
                  &  $2 \rightarrow 1$ & 219.560319 & 11.3  & 400-900  \\
${\rm HC_{3}N}$ & $9 \rightarrow 8$ & 81.881468 & 30.2 & 130-250  \\
   & $10 \rightarrow 9$ &  90.978993  & 27.2 &  130-250 \\
   & $11 \rightarrow 10$ & 100.076389 & 24.7 &  130-250 \\
%   & $23 \rightarrow 22$ & 209.230 & 11.8 &  400-1000 \\
%${\rm C}_{3}{\rm H}_{2}$ & $2_{12} \rightarrow 1_{01}$ & 85.339 & 29.0 & 130-250 \\
${\rm CH}_{3}{\rm CCH}$ & $5_{0} \rightarrow 4_{0}$ & 85.457300
& 28.9 & 130-250 \\
                                 & $6_{0} \rightarrow 5_{0}$ & 102.547984
& 24.1 & 130-250 \\ 
${\rm HCO ^{+}}$  & $1 \rightarrow 0$  &  89.188518 & 27.7  & 150-250   \\
HCN               & $1 \rightarrow 0$ & 88.631847 & 27.9 & 150-250  \\
%HNC               & $1 \rightarrow 0$ & 90.663543 & 27.3 & 150-250  \\
%${\rm H_{13}CO ^{+}}$  & $1 \rightarrow 0$  & 86.754294  & 28.5 & 150-250   \\
\hline
\end{tabular}

\begin{flushleft}
  \begin{scriptsize}
    \begin{minipage}{85mm}
      \begin{scriptsize}
Note. -- $\nu_{\rm 0}$ is the rest frequency of the line, HPBW is the ideal
half power beam width given by $1.2 \; \lambda /D$ where {\it D} is the
telescope diameter and $\lambda$ the line wavelength, $T_{\rm sys}$ is the SSB system temperature
      \end{scriptsize}
     \end{minipage}
    \end{scriptsize}
\end{flushleft}

\end{center} 
\end{table} 

%%%%%%%%%%%%%%%%%%%%%%%%%%%%%%%%%%%%%%%%%%%%%%%%%%%%%%%%%%%%%%

\section{Observations}
\label{obs}
Observations of 13 IR dark clouds were conducted at the~~IRAM 30-m telescope between July 1998 and August 1999.
Two summer runs were dedicated to spectroscopic observations, while continuum
mapping at 1.2 mm was performed during winter time. 
Line and source parameters are listed in Tables~\ref{line param} and~\ref{source param}.

%[\sout{In the following sections we give some details on the observing procedures and the data reduction techniques used.}]

\subsection{Spectroscopic observations}
Spectroscopic observations were performed during summertime under good to average weather conditions, in single-side band (SSB) mode, with receiver temperatures of about 90 K 
at 3 mm and 180 K at 1.3 mm.
The data were calibrated to the $T_{\rm A}^{*}$ scale using the chopper wheel method (Penzias \& Burrus, 1973). Discussion on the final brightness temperature scale $T'_{\rm mb}$
adopted for our data is presented in appendix~\ref{append 1.1}.
Using comparison with observations of line calibrators (Mauersberger et al. 1989),
%[\sout{, and sources from the IRAM Key-Project, thereafter IKP, Falgarone et al. 1998)}]
 we believe our absolute calibration to be better than 20\%. 
The pointing accuracy was checked by repeated continuum scans across planets and strong quasars.
The spectrometer was an autocorrelator set to a
resolution of 80 kHz, yielding velocity channels of 0.24 km s$^{-1}$ (0.1 km s$^{-1}$)
at 3 mm (1.3 mm). Additional details on the technique used to correct for baseline spurious effects in the autocorrelator are given in appendix~\ref{append 1.2}
%[\sout{A 1 MHz resolution filter bank (FB) was connected in parallel in order to accurately correct from platforming effects arising in the AC. We used the FB data to correct for the AC subband spurious effects before correcting for baseline ripples.}]

%%%%%%%%%%%%%%%%%%%%%%%%%%%%%%%%%%%%%%%%%%%%%%%%%%%%%%%%%%%%%%

\begin{table}
\begin{center}
\caption{Source parameters}
\label{source param}
\begin{small}
\begin{tabular}{l c c c} 
\hline \hline \\
Name & RA & Dec & d$^{(C)}$ \\
     & (J2000.0)   &(J2000.0)     & (kpc) \\  \hline
DF+04.36--0.06       & 17:55:53.07 & -25:13:18.7 & 3.5 \\ %/13.7\\
DF+09.86--0.04$^{(A)}$ & 18:07:37.22 & -20:25:54.5 & 2.8\\ %/13.9\\
DF+15.05+0.09$^{(A)}$ & 18:17:37.87 & -15:48:59.9 & 3.1\\ %/13.3\\
DF+18.56--0.15         & 18:25:19.52 & -12:49:57.0 & 4.0\\ %/13.3\\
DF+18.79--0.03         & 18:25:19.84 & -12:34:23.1 & 3.6\\ %/13.3\\
DF+25.90--0.17         & 18:39:10.13 & -06:19:58.8 & 5.5\\ %/13.3\\
DF+30.23--0.20$^{(B)}$ & 18:47:13.16 & -02:29:44.7 & 6.7\\ %7.9\\
DF+30.31--0.28       & 18:47:39.03 & -02:27:39.8 & 6.3\\ %/8.4\\
DF+30.36+0.11       & 18:46:21.16 & -02:14:19.0 & 5.9\\ %/8.8\\
                    &             &             & 7.4$^{(D)}$\\ %/8.8\\
DF+30.36--0.27       & 18:47:42.37 & -02:24:43.2 & 6.9\\ %/7.8\\
DF+31.03+0.27$^{(B)}$ & 18:47:00.39 & -01:34:10.0 & 4.9        \\ %/9.7-8.6\\
                      &             &             & 6.0\\ %/9.7-8.6\\
DF+36.95+0.22       & 18:57:59.51 & +03:40:33.3 & 5.0 \\%/8.5\\
DF+51.47+0.00$^{(B)}$ & 19:26:12.74 & +16:26:12.6 & 5.3$^{(D)}$\\ \hline 
\end{tabular}
\end{small}

\begin{flushleft}
  \begin{scriptsize}
    \begin{minipage}{85mm}
      \begin{scriptsize}
$^{(A)}$ Observed  in  April'99 (continuum)\\
$^{(B)}$ Observed  both in April'99 and August'99\\
$^{(C)}$ More than one value may be given in case several lines are detected on the line of sight\\
$^{(D)}$ Distance of the tangent point is assumed
      \end{scriptsize}
     \end{minipage}
    \end{scriptsize}
\end{flushleft}

\end{center} 
\end{table}

%%%%%%%%%%%%%%%%%%%%%%%%%%%%%%%%%%%%%%%%%%%%%%%%%%%%%%%%%%%%%%%%%%%%%%%%%%%

%%%%%%%%%%%%%%%%%%%%%%%%%%%%%%%%%%%%%%%%%%%%%%%%%%%%%%%%%%

 \begin{figure*}
 \begin{center}
 \vspace{140mm}
 \includegraphics{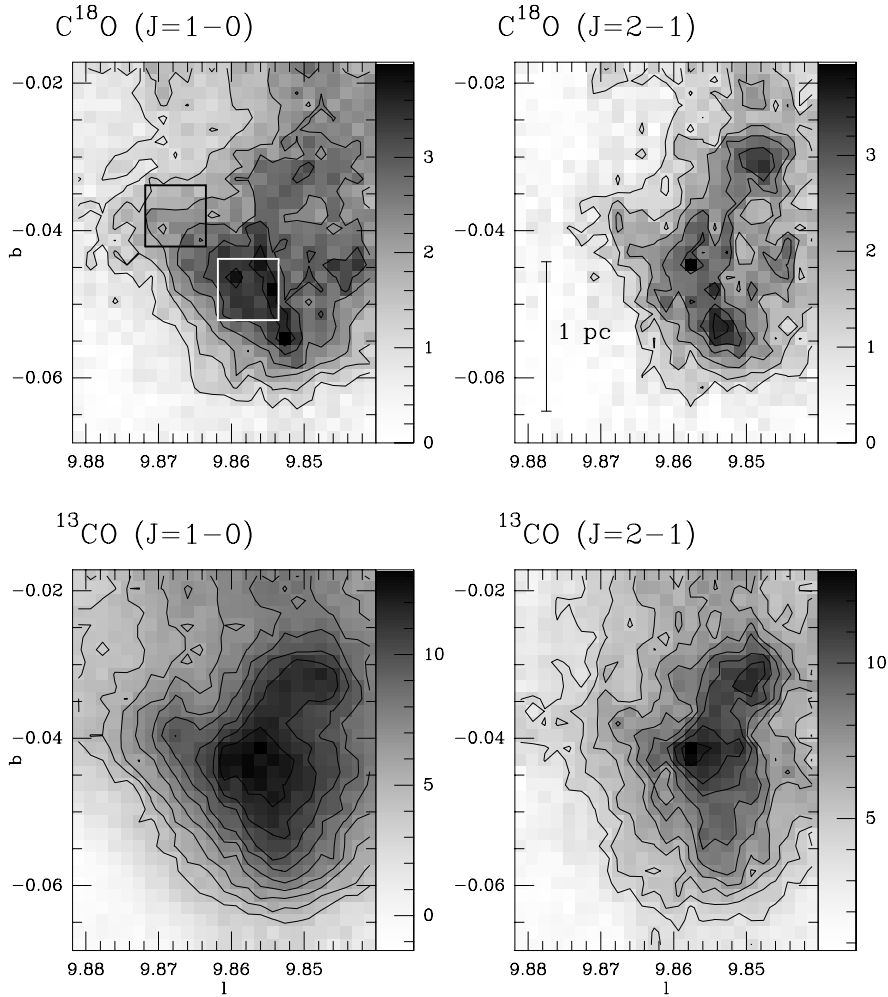}
 \caption{Maps of CO isotopomers integrated emission for DF+09.86--0.04. Contours are 1 to 4 K\,km\,s$^{-1}$ by steps of 0.5 for ${\rm C^{18}O}$ transitions and 4 to 12 K\,km\,s$^{-1}$ by steps of 1 for $^{13}{\rm CO}$ maps (${\rm T}_{\rm A}^{*}$ scale). Velocity interval is 16-20 km\,s$^{-1}$. The white box indicates the {\it core} position while the black one frames the {\it envelope} area used in the analysis.}
 \label{09 co map}
 \end{center}
 \end{figure*}

%%%%%%%%%%%%%%%%%%%%%%%%%%%%%%%%%%%%%%%%%%%%%%%%%%%%%%%%%%%

The maps were obtained with the Spectral Line On-the-Fly technique
(Ungerechts et al. 1999) with a scanning speed of 1\arcsec/second (samples every 2 seconds) and a cross-scan sampling of 6\arcsec.
%[\sout{In cross-scan direction, we adopted a {\it noise-dependent} sampling interval derived from the monitoring of the SSB system temperature. Allowing for steps of 6\arcsec~and less, we could get homogeneous fully-sampled maps. This method leads to sometimes very long mapping time under critical weather conditions.}]
We used two classes of reference positions. The targets were mapped using a close-by reference position that was then compared to a further one chosen from the
Massachusetts-Stony Brook Galactic Plane CO Survey (Sanders et al.
1986) and estimated as sufficiently free of emission.
Finally, whenever possible, map coverages in the orthogonal direction
were achieved allowing use of the PLAIT algorithm developed
by Emerson et al. (1988).

%%%%%%%%%%%%%%%%%%%%%%%%%%%%%%%%%%%%%%%%%%%%%%%%%%%%%%%%%%%%%%%%%%%%%

 \begin{figure*}
 \begin{center}
 \vspace{120mm}
 \includegraphics{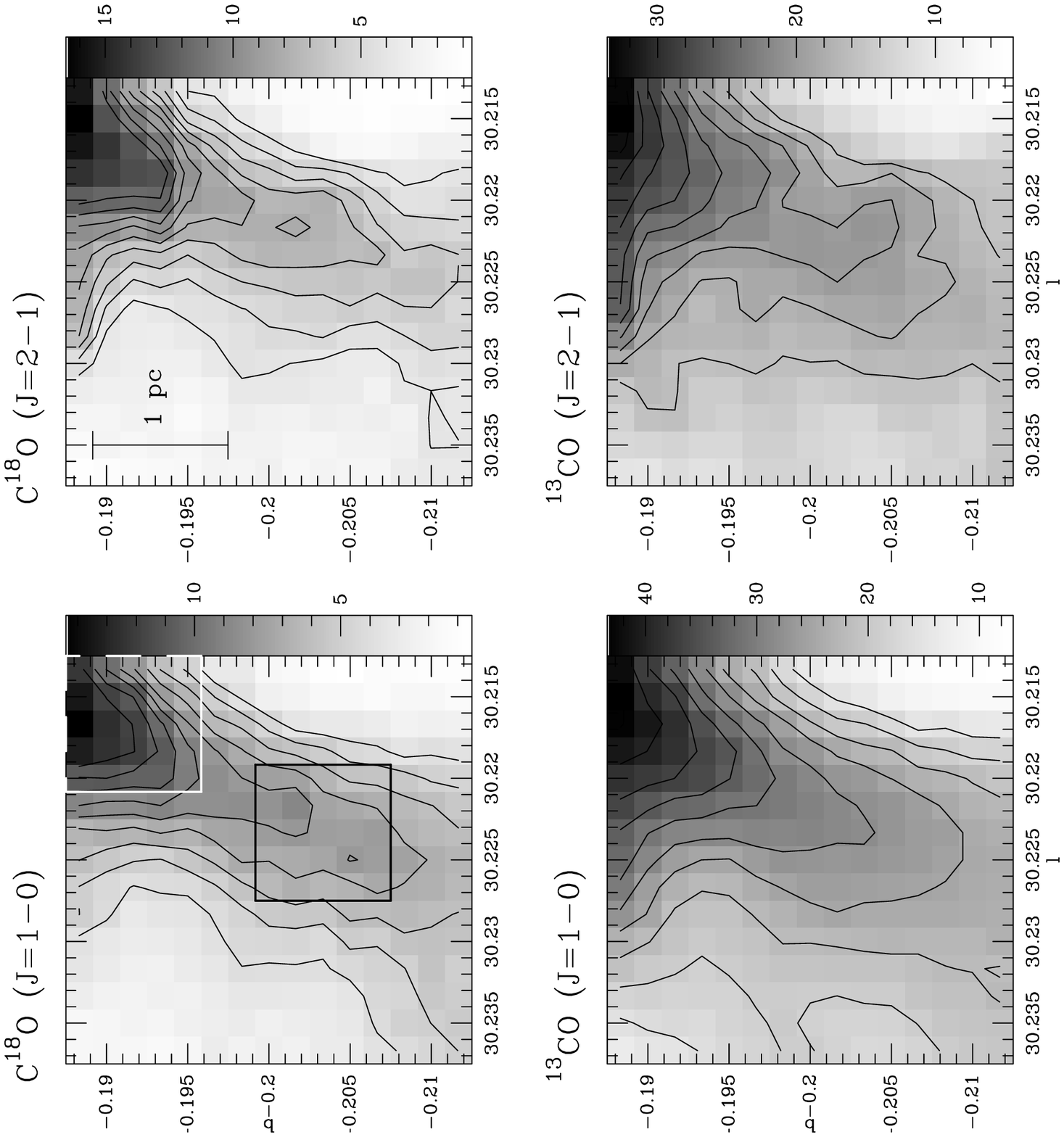}
 \caption{Same as Fig.~\ref{09 co map} for the central area of DF+30.23--0.20. Contours are 4 to 12 K\,km\,s$^{-1}$ by steps of 1 for ${\rm C^{18}O}$ transitions, 16 to 35 K\,km\,s$^{-1}$ by steps of 2 for $^{13}{\rm CO}(J=2 - 1)$ and 16 to 45 K\,km\,s$^{-1}$ by steps of 3 for $^{13}{\rm CO}(J=1 - 0)$ (${\rm T}_{\rm A}^{*}$ scale). Velocity interval is 103-107 km\,s$^{-1}$. The black box indicates the {\it filament} position while the white one frames the {\it star edge} area.} 
 \label{30c co map}
 \end{center}
 \end{figure*}

%%%%%%%%%%%%%%%%%%%%%%%%%%%%%%%%%%%%%%%%%%%%%%%%%%%%%%%%%%%%%%%%%%%%%

Maps of 3\arcmin$\times  $3\arcmin~or 4\arcmin$\times  $4\arcmin~
were obtained for 7 of the fields and narrow bands of 0.5\arcmin$\times  $3\arcmin~for the other ones. ${\rm HCO^{+}}$ and HCN $J=1-0$ lines were measured in parallel with $^{13}$CO and C$^{18}$O (${\rm HCO^{+}}$ and HCN data are not discussed here).
Examples of integrated maps are given in Fig.~\ref{09 co map} and~\ref{30c co map}.
Noise levels (rms) reached are around  0.26 K (0.53 K) at 3 mm (1.3 mm) per regridded pixel and spectral resolution elements of 80 kHz.
%[\sout{Four}] 
Three of the targets were observed in a second run in HC$_{3}$N and CH$_{3}$CCH 
(see transitions in Table~\ref{line param}). HC$_{3}$N maps of 1.4\arcmin$\times  $1.4\arcmin~
were obtained to further allow for accurate comparison of lines at different frequencies
by restoring all data to a common synthetic beam. Transitions of CH$_{3}$CCH were observed in position 
switching, yielding a 20 mK rms level per spectral resolution element.

\subsection{Continuum observations}
\label{bolo obs}
Five clouds were mapped using the MPIfR 37-channel
bolometer array centred at $\sim$\,1.2 mm (Kreysa 1992, Kramer et al. 1998). We used
the on-the-fly mode for continuum observations, consisting of
dual-beam rasters with scanning velocities of 6\arcsec/sec or
8\arcsec/sec. The spatial sampling interval in elevation was larger than generally used at the 30-m telescope. With a vertical offset of 24\arcsec~ 
(instead of 4-5\arcsec), individual detector maps are 
{\it under}-sampled but the co-added final map is fully sampled (Teyssier \& Sievers 1999).
In a relatively short time (about 20 minutes) we could cover areas of
7\arcmin$\times $6\arcmin~with homogeneous observing conditions and a small field
curvature when re-projecting in the equatorial frame.
%[\sout{To get a better signal-to-noise ratio,}] 
All maps were repeated several times at different hour angles and with different wobbler throws (46\arcsec, 56\arcsec~ and 78\arcsec) 
%[\sout{have been used}]
in order to allow the restoration of all spatial frequencies (see Emerson et al. 1995, or Pierce-Price et al. 2001). This additional information, however, is not used by the IRAM data reduction software (NIC, Brogui{\`e}re et al. 1999) that we have used to derive the images presented here. A straightforward restoration algorithm (Emerson, Klein \& Haslam 1979) is applied instead, followed by zero-order baseline substraction and a systematic skynoise removal.
Fig.~\ref{bolo maps} gives an example of two of these maps.

The beam size at 1.2 mm was measured to be $\sim$ 11\arcsec~using Uranus. The
calibration was achieved through regular on-the-fly and on-off
observations on planets (Uranus and Mars). We estimated both relative
and absolute calibration to be within $\sim$ 10\%. The zenith
atmospheric optical depth was found to be between $\sim$ 0.1 and
$\sim$ 0.3, according to regular antenna tipping.

\section{Observational properties}
\label{obs properties}

\subsection{Identification of the clouds and spatial distribution}
\begin{figure*}
 \begin{center}
 \vspace{75mm}
% \special{psfile=map_bolo.eps voffset=0 hoffset=0  hscale=60 vscale=60 angle=0}
 \includegraphics{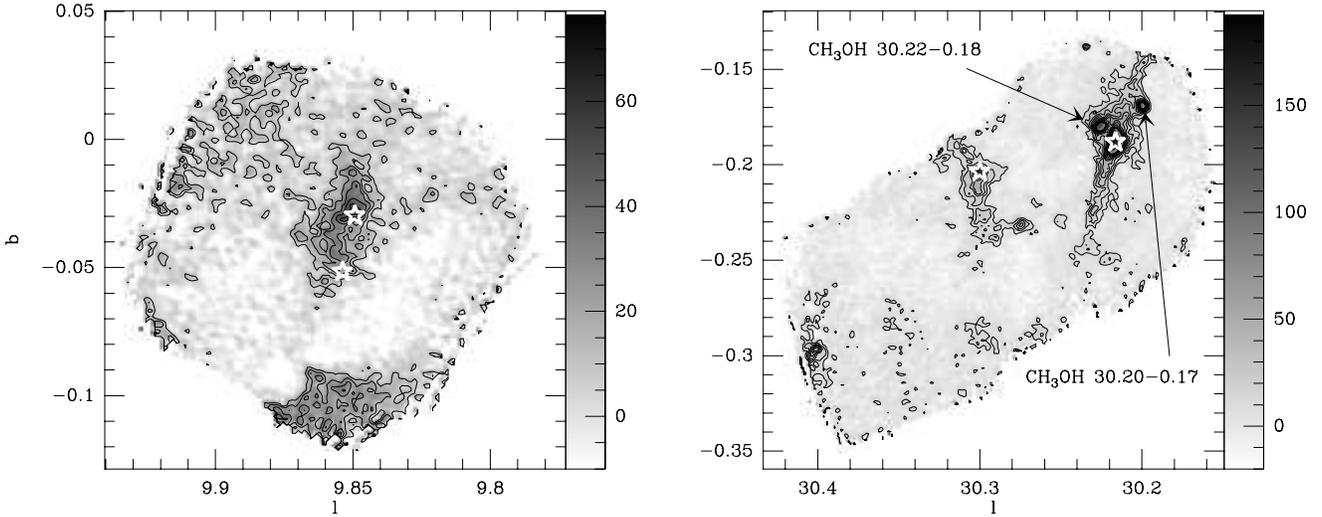}
 \caption{Continuum maps at 1.2 mm. {\it Left}: DF+09.86--0.04. Contours are 10 to 60 mJy/11\arcsec~beam by steps of 10. {\it Right}: DF+30.23--0.20 and surroundings. Contours are 10 to 100 mJy/11\arcsec~beam by steps of 10, then 100 to 200 by steps of 50. CH$_3$OH masers detected in previous surveys (Caswell et al. 1995) are indicated. White star symbols indicate embedded stars identified in the mid-IR plates. Positions are in galactic coordinates.}
 \label{bolo maps}
 \end{center}
 \end{figure*}

\begin{figure*}
 \begin{center}
 \vspace{125mm}
 \includegraphics{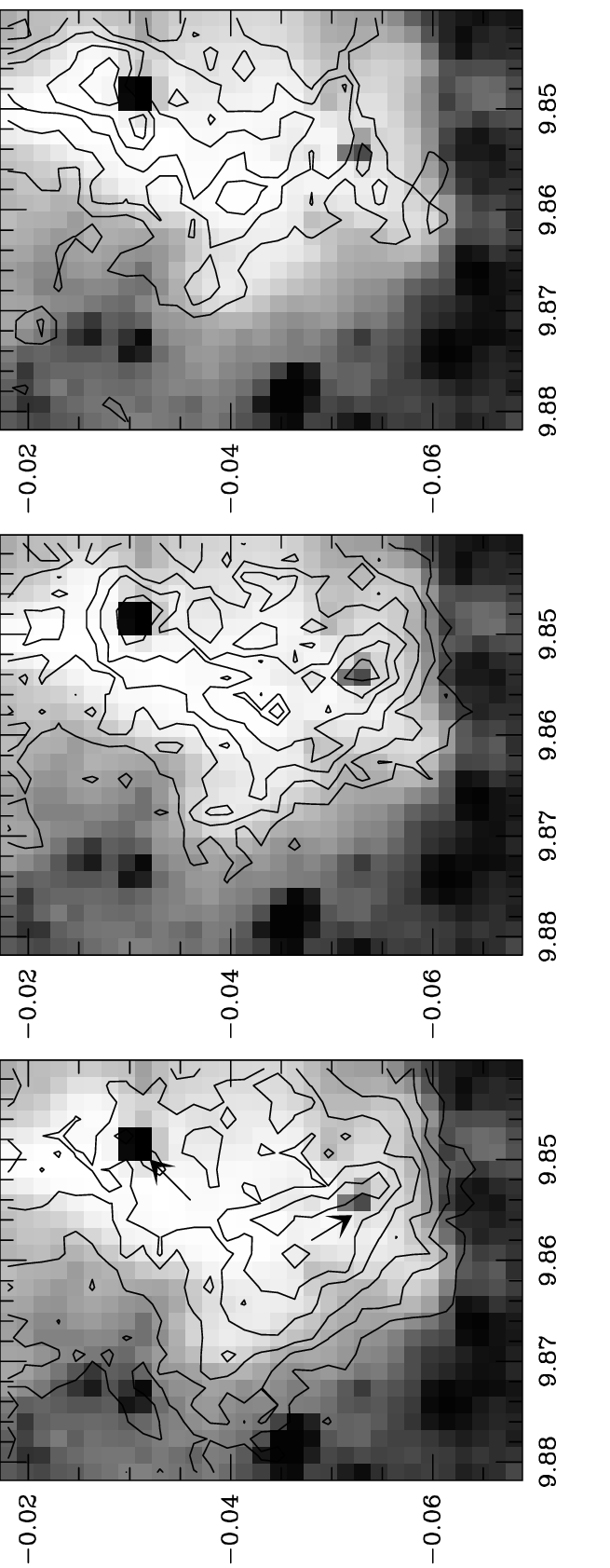}
 \includegraphics{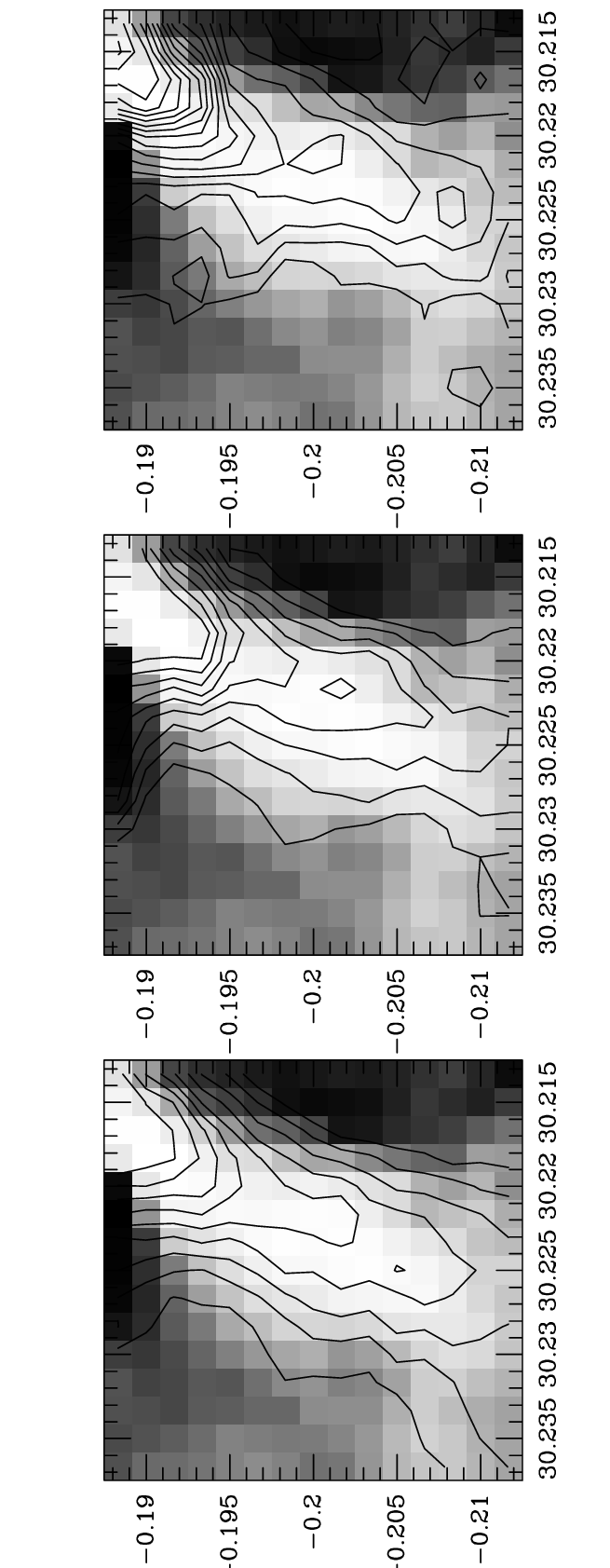}
 \caption{{\bf Upper panels}: {\it Left}: Contour map of the ${\rm C^{18}O(J=1 - 0)}$ transition integrated emission for 
DF+09.86--0.04 superimposed over the corresponding ISOGAL field in the LW3 (15 $\mu$m) filter. ISOGAL data have been reversed so that absorption features appear with positive values (see text). Arrows indicate positions of embedded stars. {\it Middle}: same for the ${\rm C^{18}O(J=2 - 1)}$ transition. In both cases, contours are 1 K\,km\,s$^{-1}$ to 4 K\,km\,s$^{-1}$ by steps of 0.5 (${\rm T}_{\rm A}^{*}$ scale), velocity interval is 16 to 20 km\,s$^{-1}$.
{\it Right}: Contour map of 1.2 mm continuum emission for 
DF+09.86--0.04 superimposed over the corresponding ISOGAL field in the LW3 filter. Contours are 10 to 60 mJy/11\arcsec~beam by steps of 10.
{\bf Lower panels}: same as upper panels, for DF+30.23--0.20. Contours are 4 to 12 K\,km\,s$^{-1}$ by steps of 1 for both C$^{18}$O maps, and 10 to 170 mJy/11\arcsec~beam by steps of 10 for the continuum map.
Coordinates are galactic longitude and latitude.}
 \label{co/iso map}
 \end{center}
 \end{figure*}

The infrared absorption features coincide remarkably with the C$^{18}$O emission ridges (left and middle panels of Fig.~\ref{co/iso map}), as well as with the continuum emission, although sometimes to a smaller extent (right hand panels of Fig.~\ref{co/iso map}).
%[\sout{All $^{13}$CO and C$^{18}$O maps present remarkable correlations with the infrared absorption (left and middle panels of Fig.~\ref{co/iso map}), with slight differences depending on the filter used by ISOGAL (see Hennebelle et al, 2001). This correlation is also observed in the continuum data although sometimes to a smaller extent (right hand panels of Fig.~\ref{co/iso map}). Regarding this correlation issue, one should not forget that the bolometer probes the entire line of sight while spectroscopy focuses on a given velocity component and mid-IR absorption is biased towards closer dense structures without strong heating sources. Nevertheless, the structures seen in all cases can undoubtfully be associated, sometimes combining several velocity components. Still, their interpretation remains complex, as is inescapable in the inner Galaxy.}]
In some of the clouds local peaks in the millimetre maps coincide with the mid-IR emission of embedded stars or stellar clusters seen by ISO. The identification of OH and CH$_3$OH masers (Caswell et al. 1995, see Fig.~\ref{bolo maps}) associated with these bright sources give additional evidence that stars may have formed recently.
%[\sout{seen in emission in the CO isotopomer transitions, proving that these stars indeed belong to the mid-IR absorption structures since they are detected at the same velocity as the envelopes tracing their surrounding.}]

Fig.~\ref{bolo maps} and~\ref{co/iso map} illustrate such cases. For DF+09.86--0.04, the embedded stars appear as two small black squares indicated by arrows in the infrared data (marked with white stars in the continuum map).
%[\sout{(images were reversed so that the absorption features have positive values)}]
For DF+30.23--0.20, the stars themselves have not been mapped in CO, but are clearly seen as strong spots in continuum (white stars symbols).
%[\sout{due to their heating effect on the filament, a strong temperature gradient is observed.}]
The dark filament turns to a bright filament in the mid-IR, while its millimetre emission is strongly reinforced around newly formed stars (Fig.~\ref{bolo maps}).
%[\sout{As expected, these objects present different physical properties from dark clouds lacking significant heating sources (see Sect.~\ref{properties}).}]

%[\sout{In several cases, OH and CH$_3$OH masers (Caswell et al. 1995) are associated with bright sources seen in continuum and in the CO isotopomers, suggesting that star formation may have already started in some of our clouds.}]
 
%[\sout{In some rare cases the emission also arises from the close environment of stars seen in emission in the mid-IR data, tracing different physical conditions.}]

%[\sout{The shapes range from globules (Fig.~\ref{09 co map}) to filaments (Fig.~\ref{30c co map}), with corresponding typical sizes varying between 5 and 1 pc for the smallest structures.}]

\begin{figure*}
 \begin{center}
% \vspace{75mm}
 \centerline{\epsfig{file=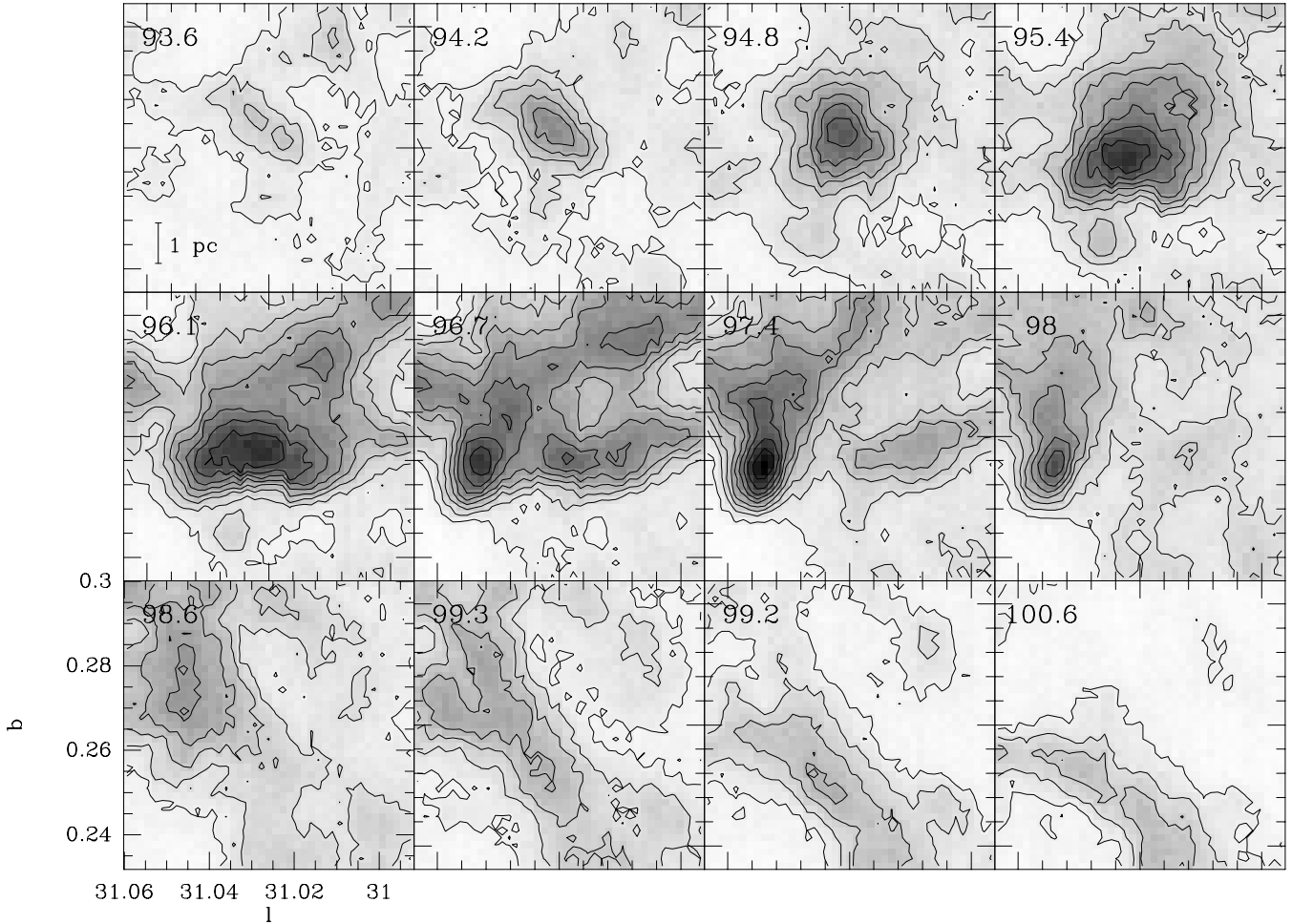, height=17.7cm, angle = 270}}
% \special{psfile=vel_map_31.ps voffset=-410 hoffset=0  hscale=80 vscale=80 ang%le=0}
 \caption{$^{13}{\rm CO}(J=1 - 0)$ channel maps of DF+31.03+0.27 averaged over 3 channels of 0.21 $\rm km~s^{-1}$. The central velocity in km s$^{-1}$ is indicated on the upper left corner for each map. Contours are 0.4 to 6.7 $\rm K~km~s^{-1}$ by steps of 0.7 ($T^*_{\rm A}$ scale).}
 \label{vel map 31}
 \end{center}
 \end{figure*}

Some of the clouds present several emission line components along the line of sight (e.g. DF+30.36+0.11 and DF+31.03+0.27, which are the most distant clouds of our sample). The structures associated with different velocity components present very different shapes and orientations, which sometimes need to be combined to reproduce the shape of the dark cloud. For all clouds the velocity field exhibits significant structure, as illustrated for one component of the DF+31.03+0.27 cloud by the channel maps of Fig.~\ref{vel map 31}.
In this object, the profiles are complex and the gas structures seem to be connected over a wide velocity range.

%%%%%%%%%%%%%%%%%%%%%%%%%%%%%%%%%%%%%%%%%%%%%%%%%%%%%%%%%%%%%%%%%%%

%%%%%%%%%%%%%%%%%%%%%%%%%%%%%%%%%%%%%%%%%%%%%%%%%%%%%%%%%%%%%%%%

%%%%%%%%%%%%%%%%%%%%%%%%%%%%%%%%%%%%%%%%%%%%%%%%%%%%%%%%%%%%%%%%%%%%%%%%%%%%%%%

% \begin{figure*}
% \begin{center}
% \vspace{75mm}
% \special{psfile=vel_map_09.ps voffset=-410 hoffset=0  hscale=80 vscale=80 ang%le=0}
% \centerline{\epsfig{file=vel_map_09.ps, width=13.cm, angle = 270}}
% \caption{$^{13}{\rm CO}(J=1 - 0)$ channel maps of DF+09.86--0.04 averaged over 3 contiguous channels of 0.21 $\rm km~s^{-1}$ each. The central velocity is indicated on the upper left corner for each map. Contours are 0.3 to 4 $\rm K~km~s^{-1}$ by steps of 0.4 ($T^*_{\rm A}$ scale).}
% \label{vel map 09}
% \end{center}
% \end{figure*}

%%%%%%%%%%%%%%%%%%%%%%%%%%%%%%%%%%%%%%%%%%%%%%%%%%%%%%%%%%%%%%%%%%%%%

%%%%%%%%%%%%%%%%%%%%%%%%%%%%%%%%%%%%%%%%%%%%%%%%%%%%%%%%%%%%%%%%

 \begin{figure*}
 \begin{center}
 \vspace{75mm}
 \includegraphics{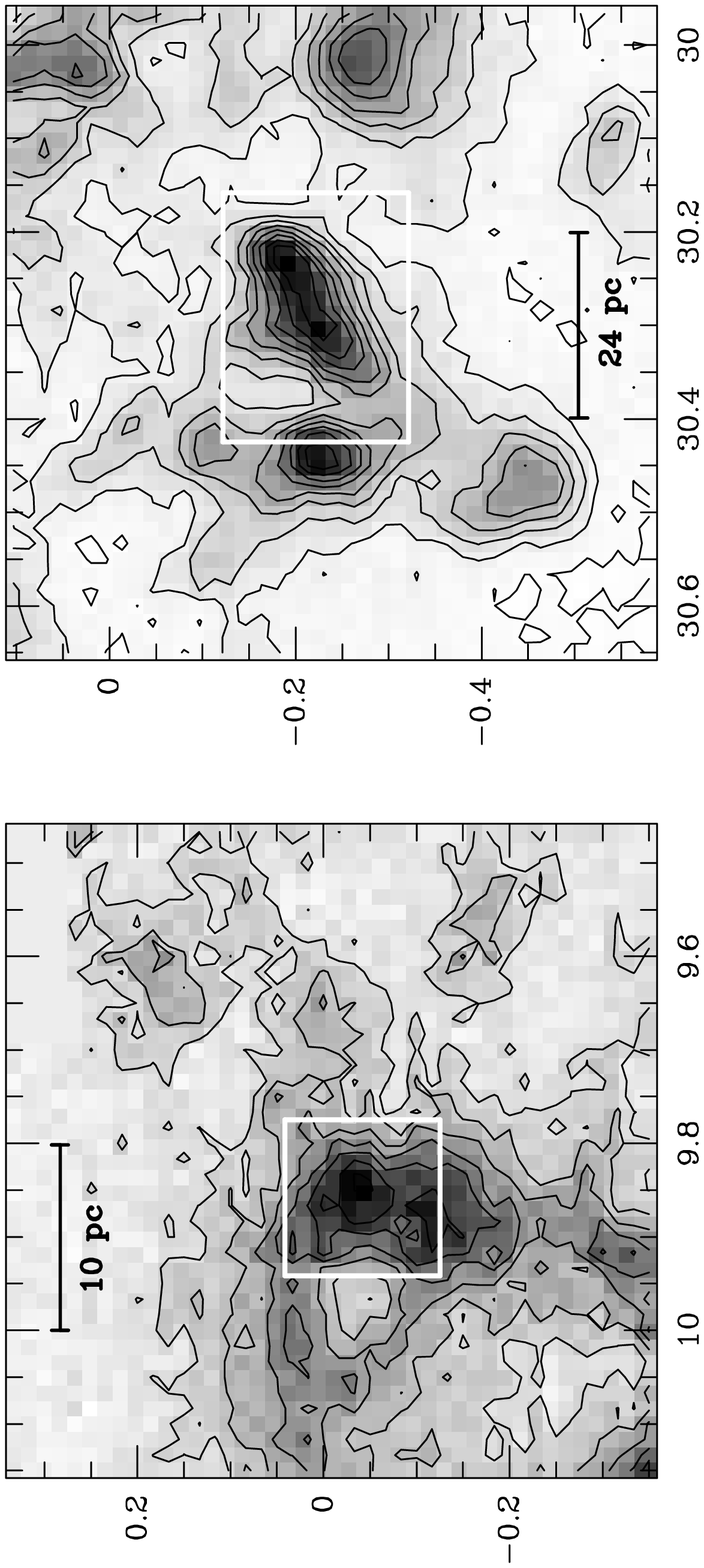}
 \caption{Large field maps of DF+09.86--0.04 (left) and DF+30.23--0.20 (right) in the $^{13}{\rm CO}(J=1 - 0)$ transition. Beam HPBW is 2.5\arcmin~ and sampling interval is 1\arcmin. Velocities are integrated between 16 and 20 km~s$^{-1}$ and 104 and 107 km~s$^{-1}$ respectively. The spatial connections with larger scale structures is clear. White boxes indicate the extent of the bolometer maps displayed in Fig.~\ref{bolo maps}.}
 \label{co large maps}
 \end{center}
 \end{figure*}

%%%%%%%%%%%%%%%%%%%%%%%%%%%%%%%%%%%%%%%%%%%%%%%%%%%%%%%%%%%%%%%%%%%%%

Large field maps obtained with the 4-m Nanten telescope in $^{13}$CO$(J=1-0)$ (Zagury et al., yet unpublished data) and shown
on Fig.~\ref{co large maps} for DF+09.86--0.04 and DF+30.23--0.20 indicate that the IR dark clouds are not isolated objects: they are embedded in quiescent GMC's, most of them belonging to the cold GMC population detected earlier in CO galactic surveys (e.g. Sanders et al. 1986).
%[\sout{This is illustrated on Fig.~\ref{co large maps} for DF+09.86--0.04 and DF+30.23--0.20 (see captions for details).}]
This contradicts Egan et al.'s (1998) claim that the IRDCs are isolated objects.

%[\sout{A quantitative analysis of the line intensities is given in Sect.~\ref{profile analysis} separately for the emission peaks (core) and periphery (envelope), and for the line centres and wings.}]

Using the galactic rotation model of Burton et al. (1991), we are able to derive  kinematic distances  of these objects from the Doppler shift of the line. Assuming that the kinematic distance ambiguity is solved by the absorption bias in favour of the nearest position, we derive distances between $2.8$ and $7.4$ kpc (see Table~\ref{source param}). 
%[\sout{When the object velocity was greater than the tangent point velocity, we assumed the distance to be the one at the tangent point.}]
For all observed clouds the inferred kinematic distances  are consistent with objects lying within or in front of the molecular ring.

%[\sout{\subsection{Spatial correlations}}]

%[\sout{\subsection{Spatial and velocity distribution} \label{spatial distrib} The maps revealed a broad variety of properties both in the spatial and in the spectral dimensions.}]

%[\sout{\subsubsection{Temperature scales} \label{T scale}}]

%%%%%%%%%%%%%%%%%%%%%%%%%%%%%%%%%%%%%%%%%%%%%%%%%%%%%

 \begin{figure*}
 \begin{center}
 \vspace{75mm}
 \includegraphics{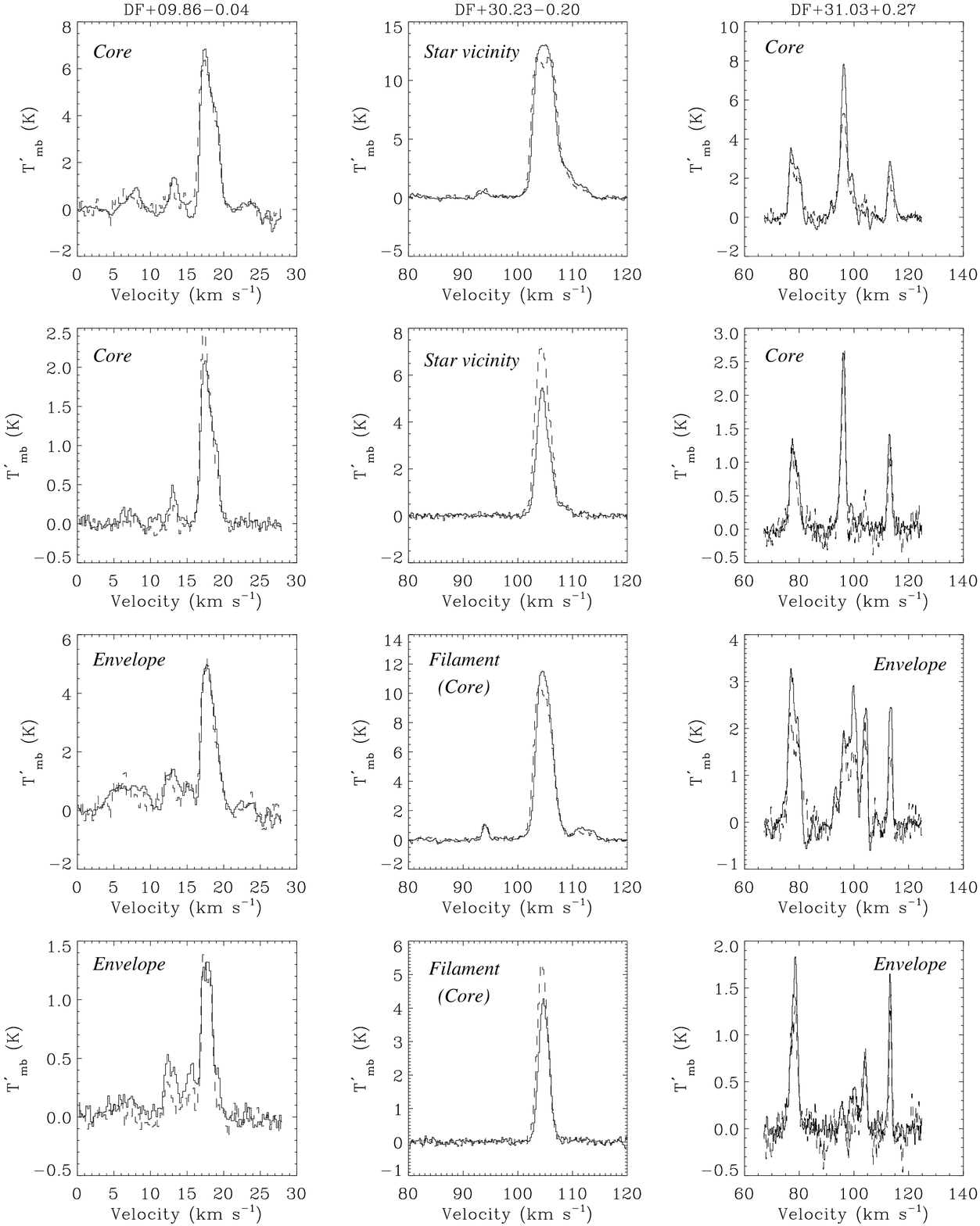}
\kern145mm\vbox{ \caption{Average spectra of CO isotopomers for DF+09.86--0.04, DF+30.23--0.20 and DF+31.03+0.27. First row of panels: $^{13}{\rm CO}(J=1 - 0)$ (full lines) and $^{13}{\rm CO}(J=2 - 1)$ (dashed lines) transitions in the core area. Second row of panels: ${\rm C^{18}O(J=1 - 0})$ (full lines) and  ${\rm C^{18}O(J=2 - 1})$ (dashed lines) in the {\it core} (or {\it star vicinity}) area. Third and fourth rows of panels: same as 1st and 2nd rows for a different area of the same clouds. The spectra have been convolved to identical resolution (22\arcsec and 0.213 km~s$^{-1}$ channels), and averaged over 30\arcsec~boxes. Temperatures are in the $T'_{\rm mb}$ scale (see appendix~A.1).}}
 \end{center}
 \label{spectra table}
 \end{figure*}

%%%%%%%%%%%%%%%%%%%%%%%%%%%%%%%%%%%%%%%%%%%%%%%%%%%%%%%%

\begin{table*}%[l]
\begin{center}
\caption{CO line parameters for spectra averaged over 30\arcsec$\times$ 30\arcsec areas}
\label{CO table}
\begin{scriptsize}
\begin{tabular}{l c c c c c c c c c} 
\hline \hline \\
 & & \multicolumn{2}{c}{${\rm C^{18}O}$ $^{(B)}$} & \multicolumn{2}{c}{$^{13}{\rm CO}$ $^{(B)}$} & \multicolumn{4}{c}{$^{(C)}$}\\
Name & $v_0$ $^{(A)}$ &  $(J=1 \rightarrow 0)$ &  $(J=2 \rightarrow 1)$ & $(J=1 \rightarrow 0)$ &  $(J=2 \rightarrow 1)$ & $R_{18}$ & $R_{13}$ & $R_{10}$  &  $R_{21}$\\
   & &  & & & & & & \\  \hline
DF+04.36--0.06           &  &  &  & & & & &\\ 
{\it core}      & 11.42 & 3.34(0.08) & 2.26(0.17)&   --      & 5.00(0.14) & 0.677 & --    &   --  & 2.21  \\ 
 			& 	& 1.71 	     & 1.78  	 &           & 1.80 	  &(0.065)&       &       &(0.23) \\ 
{\it envelope}  & 11.37 & 2.10(0.08) & 1.54(0.17)&  --	     & 4.05(0.14) & 0.731 &  --   &  --   & 2.64  \\ 
			&  	& 1.61 	     & 1.76 	 & 	     & 2.70 	  &(0.106)&       &       &(0.38) \\ 

DF+09.86--0.04  &  &  &  & & & & &\\ 
{\it core$^{(E)}$}      & 17.79 &  1.97(0.05)& 2.29(0.10)& 6.90(0.05)&	6.22(0.10)& 1.160 &  0.900& 3.50  & 2.72  \\ 
			&  	&  2.01      & 1.81 	 & 2.52      & 	2.80	  &(0.081)&(0.020)&(0.11) &(0.16) \\ 
{\it envelope}          & 17.69 &  1.30(0.05)& 1.39(0.10)& 4.94(0.05)& 4.87(0.10) & 1.065 &  0.986& 3.78  & 3.50  \\ 
			&  	&  2.12      & 1.63 	 & 2.26      & 2.23 	  &(0.118)&(0.029)&(0.19) &(0.32) \\ 

DF+15.05+0.09 &  &  &  & & & & &\\ 
{\it core$^{(E)}$}      & 29.95 & 2.84(0.08) & 2.16(0.13) & 5.80(0.08)&	4.16(0.11)& 0.763 & 0.717 & 2.04  & 1.92  \\ 
			&  	& 2.05 	     & 2.13  	  & 2.56     & 1.70 	  &(0.066)&(0.029)&(0.09) &(0.17) \\
{\it envelope}          & 29.97 & 2.02(0.08) & 1.34(0.13) & --       &	3.27(0.11)& 0.664 & --    & --    & 2.44  \\ 
			&  	& 2.13 	     & 2.20  	  &          & 1.80 	  &(0.091)&       &       &(0.32) \\

DF+18.56--0.15 &  &  &  & & & & &\\ 
{\it core}              & 50.50 & 3.11(0.09) & 1.71(0.17) &6.86(0.09)&	4.54(0.16)& 0.548 & 0.662 & 2.20  & 2.66  \\ 
			&  	& 1.90 	     & 1.40  	  & 3.25     & 3.31 	  &(0.070)&(0.032)&(0.09)&(0.36) \\
{\it envelope}          & 50.60 & 1.88(0.09) & 1.16(0.17) &5.33(0.09)&	3.32(0.16)& 0.621 & 0.622 & 2.85  & 2.85  \\ 
			&  	& 1.53 	     & 1.27  	  & 2.96     & 2.84 	  &(0.120)&(0.041)&(0.19)&(0.55) \\

DF+30.23--0.20 &  &  &  & & & & &\\ 
{\it star vicinity$^{(E)}$} & 104.7 &  5.53(0.05)& 7.62(0.13)&14.40(0.05) & 13.09$^{(D)}$(0.13)& 1.378 &  0.912& 2.60  & 1.72  \\
			&  	&  2.52      & 2.77 	 & 3.88	     & 	4.82	  &(0.037)&(0.012)&(0.03) &(0.05) \\
{\it filament}          & 104.7 &  4.23(0.05)& 5.30(0.13)& 11.90(0.05)&10.87 (0.13)& 1.252 &  0.914& 2.81  & 2.05  \\
			&       &  2.23      & 2.28  	 & 3.22	     & 	3.60	  &(0.047)&(0.014)&(0.05) &(0.08) \\

DF+30.36+0.11 &  &  &  & & & & &\\ 
{\it core$^{(E)}$}      & 96.1  &  2.40(0.09)& 3.23(0.17)& 8.32(0.09)&  7.53(0.17)& 1.349 &0.905  & 3.47  & 2.33  \\
(1$^{st}$ component)	&       &  3.64      & 3.07 	 & 3.42      & 	3.01	  &(0.122)&(0.030) &(0.17) &(0.18) \\
{\it envelope}          & 96.0  &  2.29(0.09)& 2.28(0.17)& --        &  7.23(0.17)& 0.997 & --    & --    & 3.17  \\
 			&  	&  3.14      & 2.724	 &           & 	4.09	  &(0.114)&       &       &(0.31) \\

{\it core$^{(E)}$}      & 111.0 &  4.00(0.09)& 4.77(0.20)& --        &  9.47(0.18)& 1.191 & --    &   --  & 1.99  \\
(2$^{nd}$ component) 	&       &  1.27      & 1.31 	 &           & 	1.72	  &(0.076)&       &       &(0.12) \\
{\it envelope}          & 111.3 &  1.40(0.09)& 1.94(0.20)& 5.65(0.09)&	4.52(0.18)& 1.383 & 0.800 &4.04   & 2.33  \\
			&  	&  1.57      & 1.77 	 &  2.22     & 	2.22	  &(0.230)&(0.045)&(0.32) &(0.33) \\

DF+31.03+0.27 &  &  &  & & & & &\\ 
{\it core}              & 77.8  &  2.12(0.05)& 1.56(0.13)&4.76(0.05) &	2.85(0.13)& 0.736 &  0.597& 2.25  & 1.82  \\
(1$^{st}$ component)	&       &  2.81      & 2.29 	 & 3.78	     & 	2.98	  &(0.078)&(0.033)&(0.08) &(0.23) \\
{\it envelope}          & 78.2  &  1.36(0.05)& 0.86(0.13)& 3.65(0.05)&	2.35(0.13)& 0.633 &  0.645& 2.69  & 2.73  \\
 			&  	&  3.17      & 2.86 	 & 3.81	     & 	3.82	  &(0.118)&(0.044)&(0.14) &(0.55) \\

{\it core}              & 96.0  &  2.78(0.05)& 2.96(0.17)& 8.25(0.05)&	5.38(0.15)& 1.063 &  0.653& 2.96  & 1.82  \\
(2$^{nd}$ component) 	&  	&  2.05      & 1.82 	 & 3.42	     & 	3.10	  &(0.080)&(0.022)&(0.07) &(0.15) \\
{\it envelope}          & 96.2  &  1.51(0.05)& 1.56(0.17)& 5.73(0.05)&	4.18(0.15)& 1.033 &  0.729& 3.78  & 2.67  \\
			&  	&  3.71      & 2.49 	 & 4.00	     & 	3.69	  &(0.146)&(0.033)&(0.16)&(0.38) \\

DF+51.47+0.00 &  &  &  & & & & &\\ 
 {\it core}             & 54.74 &  2.54(0.04)& 2.05(0.13)& 4.82(0.04)&	3.27(0.12)& 0.806 &  0.678& 1.90  & 1.60  \\
			&  	&  1.81      & 1.80 	 & 3.40	     & 	2.91	  &(0.063)&(0.030)&(0.04) &(0.16) \\
{\it envelope}          & 54.38 &  1.73(0.04)& 1.64(0.13)& 4.51(0.04)&	3.22(0.12)& 0.948 &  0.714& 2.61  & 1.96  \\
			&  	&  2.12      & 2.06 	 & 3.31	     & 	2.82	  &(0.097)&(0.032)&(0.08) &(0.23) \\ \hline
\end{tabular}
\end{scriptsize}
\begin{flushleft}
\begin{scriptsize}
\begin{tabular}{l}
$^{(A)}$ In km\,s$^{-1}$. Centroid velocities are nearly the same for all transitions. We give here the velocity of the ${\rm C^{18}O(J=1 - 0})$ line.\\
$^{(B)}$ In each case, we give the peak temperature (K, in ${\rm T'_{mb}}$ scale, see appendix A.1) with its noise r.m.s. ($1 \sigma_{rms}$) in brackets and the line width \\(km\,s$^{-1}$, FWHM=2.35$\sigma$). Temperatures correspond to data smoothed to the same spatial (22\arcsec) and velocity (0.213 km~s$^{-1}$) resolution.\\
$^{(C)}$ Brackets give the uncertainty on ratios based only on the spectra r.m.s. noise.\\
$^{(D)}$ Extrapolated from self-reversed spectra.\\
$^{(E)}$ Overlap with star clusters identified on the ISOGAL plates
\end{tabular}
\end{scriptsize}
\end{flushleft}
\end{center} 
\end{table*}

%%%%%%%%%%%%%%%%%%%%%%%%%%%%%%%%%%%%%%%%%%%%%%%%%%%%%%%%%%%%%%%%%%%%%%%%%%%%%%%

 \begin{figure*}
 \begin{center}
 \centerline{\epsfig{file=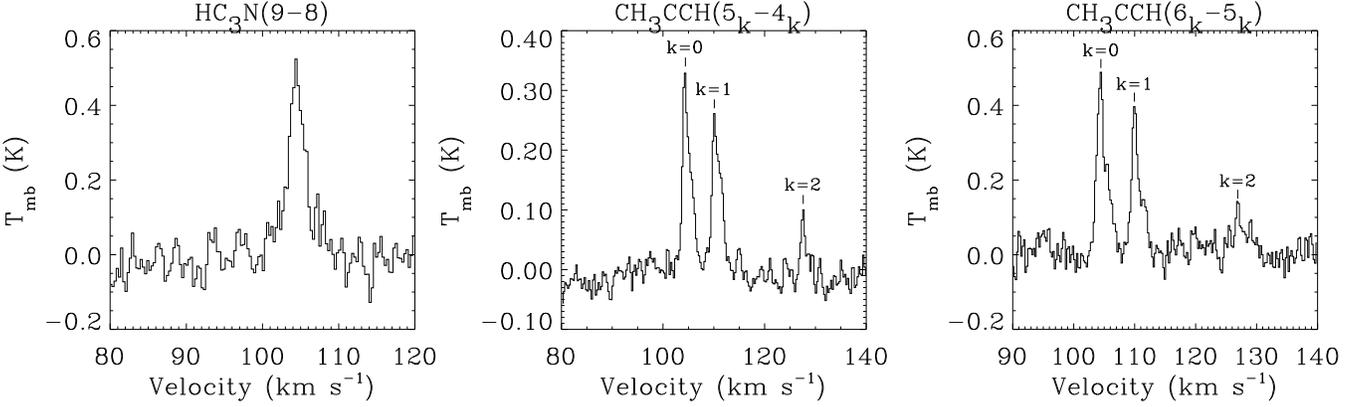, height=5.4cm}}
 \caption{Sample spectra of HC$_3$N, CH$_3$CCH($J=5-4$) and CH$_3$CCH($J=6-5$) for DF+30.23--0.20 in the star vicinity.}
 \label{hc3n spec}
 \end{center}
 \end{figure*}

%%%%%%%%%%%%%%%%%%%%%%%%%%%%%%%%%%%%%%%%%%%%%%%%%%%%%%%%%%%%%%%%%%%%

\subsection{Line intensities and velocity structure}
\label{profiles}
Figs.~7 and~\ref{hc3n spec} give examples of the line profiles observed in three of the clouds for so-called {\it core} (C$^{18}$O emission peaks) and {\it envelope} (core periphery) areas. The lines, although often centrally peaked, rarely exhibit symmetric Gaussian profiles. 
%[\sout{The more or less pronounced red or blue wings indicate that macro-turbulence may play a significant role in these clouds.}]
%[\sout{Nevertheless, for our analysis, we used the Gaussian fit elements when reliable enough. In other cases, the central velocity and the line width were deduced from the line moments.}]
In the 2 cases where embedded stars are associated with the clouds, the profiles appear as self-reversed in $^{13}$CO$(J=2-1)$ and flat-topped in $^{13}$CO$(J=1-0)$. DF+30.23--0.20 illustrates one of these cases (upper panel of the second column of Fig.~7). The line broadening due to a higher opacity 
%[\sout{and velocity dispersion}]
in the star vicinity is quite significant.

%[\sout{The separation between cores and envelopes has direct consequences on the volume of gas traced by each molecule. In particular, the HC$_{3}$N emission is only observed in the core area where the density is large enough. CO isotopomers on the other hand are also detected in envelope regions.}]

%[\sout{Regarding}]
The absolute line intensities
%[\sout{the values derived}]
are smaller than what would be expected from such massive dense clouds if extrapolated from values observed in local dark clouds (Table~\ref{CO table}). This effect is even stronger for weaker lines like HC$_3$N or CH$_3$CCH where the lines are a factor of 5 weaker than in e.g. TMC-1 (Pratap et al. 1997). 
Part of this weakness may be attributed to a very low kinetic temperature or a small filling factor. But chemical effects or, most likely, depletion onto grains should be also considered. On the other hand, the lines are broader than in local dark clouds: this is not surprising given the large size and mass of the objects.

%%%%%%%%%%%%%%%%%%%%%%%%%%%%%%%%%%%%%%%%%%%%%%%%%%%%%%%%%%%%%%%%%%%%%

\begin{figure*}
 \resizebox{12cm}{!}{\includegraphics{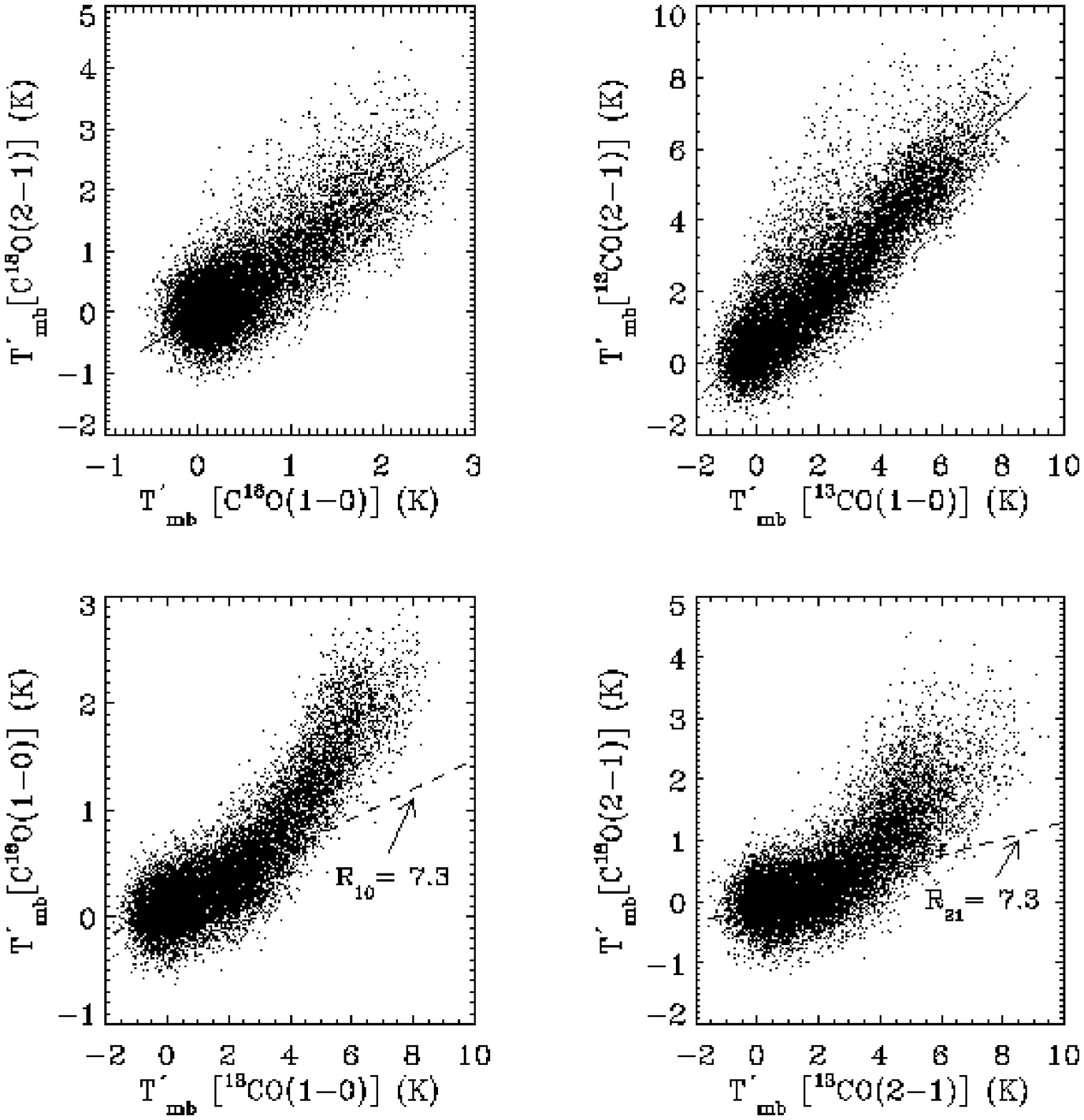}}
 \hfill
 \parbox[b]{55mm}{
   \caption{Scatter plots of CO isotopomer pairs of temperatures ${\rm T'_{mb}}$ for DF+09.86--0.04 (see text). The data have been smoothed to a 22\arcsec~ spatial and 0.2 km s$^{-1}$ spectral resolution. Velocities considered are between 16 and 20 km\,s$^{-1}$. Each 3-d pixel is represented by a dot. Best-fit results are plotted for the upper panels. Local ISM ratios are indicated by dashed-lines in the bottom diagrams.}
   \label{09 scatt}}
 \end{figure*}

%%%%%%%%%%%%%%%%%%%%%%%%%%%%%%%%%%%%%%%%%%%%%%%%%%%%%%%%%%%%%%%%%%%%%

 \begin{figure*}
% \resizebox{12cm}{!}{\includegraphics{scatt_30c.ps}}
 \resizebox{18cm}{!}{\includegraphics{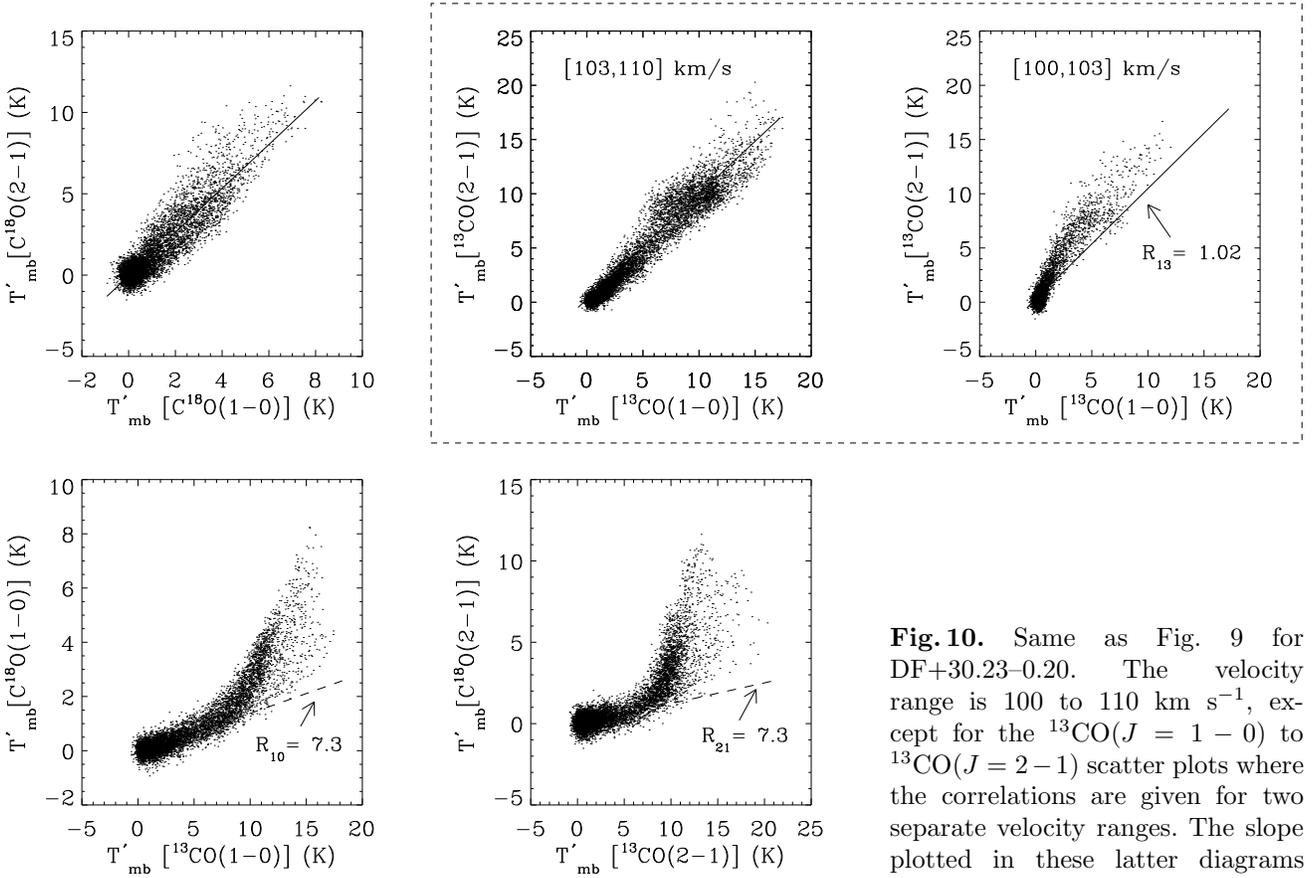}}
 \vfill
 \kern-38mm\leavevmode\kern125mm\parbox[b]{55mm}{ \caption{Same as Fig.~\ref{09 scatt} for DF+30.23--0.20. The velocity range is 100 to 110 $\rm km~s^{-1}$, except for the $^{13}{\rm CO}(J=1 - 0)$ to $^{13}{\rm CO}(J=2-1)$ scatter plots where the correlations are given for two separate velocity ranges. The slope plotted in these latter diagrams corresponds to the best-fit results of the data considered between 103 and 110 $\rm km~s^{-1}$.}
 \label{30 scatt}}
 \end{figure*}

%%%%%%%%%%%%%%%%%%%%%%%%%%%%%%%%%%%%%%%%%%%%%%%%%%%%%%%%%%%%%%%%%%%%%

%%%%%%%%%%%%%%%%%%%%%%%%%%%%%%%%%%%%%%%%%%%%%%%%%%%%%%%%%%%%%%

\begin{table}
\begin{center}
\caption{$J=2-1$ to $J=1-0$ ratios for C$^{18}$O and $^{13}$CO estimated over the entire clouds.}
\label{ratio}
\begin{small}
\begin{tabular}{l c c c c} 
\hline \hline \\
Name & $R_{18}$ & $R^I_{18}$$^{(A)}$ & $R_{13}$ & $R^I_{13}$$^{(A)}$ \\
 \hline
DF+04.36$-$0.06       & 0.66 & 0.84 & 0.59 & 0.60\\
DF+09.86$-$0.04       & 0.96 & 1.26 & 0.93 & 0.92\\
DF+15.05+0.09	      & 0.62 & 0.82 & 0.72 & 0.78\\
DF+18.56$-$0.15       & $-^{(B)}$    & $-^{(B)}$    & 0.59 & 0.61\\
DF+18.79$-$0.03       & 1.02 & 1.18 & 0.66 & 0.70\\
DF+25.90$-$0.17       & $-^{(B)}$    & 0.70 & 0.53 & 0.60\\
DF+30.23$-$0.20$^{(C)}$	      & 1.58 & 1.63 & 1.02 & 1.01\\ 
DF+30.36+0.11         & 1.07 & 1.17 & 0.74 & 0.87\\
~``~(2$^{nd}$ component)  & 0.90 & 1.24 & 0.94 & 0.79\\
DF+31.03+0.27         & 0.63 & 0.85 & 0.61 & 0.69\\
~``~(2$^{nd}$ component)  & 0.84 & 1.07 & 0.75 & 0.82\\
DF+51.47+0.00         & 0.70 & 0.78 & 0.60 & 0.63\\ \hline
Mean			&0.90 & 1.05 & 0.72 & 0.75 \\
(1-$\sigma$ deviations)			&${\pm0.28}$ & ${\pm 0.27}$ & ${\pm 0.15}$ & ${\pm 0.13}$ \\ \hline
\end{tabular}
\end{small}

\end{center} 
\begin{scriptsize}
\kern-2mm $^{(A)}$ Ratio of integrated intensities\\
$^{(B)}$ Dispersion is too large\\
$^{(C)}$ Measured between 103 and 110 km~s$^{-1}$.
\end{scriptsize}

\end{table}

%%%%%%%%%%%%%%%%%%%%%%%%%%%%%%%%%%%%%%%%%%%%%%%%%%%%%%%%%%%%%%%%%%%%%%%%%%%1

Line intensities ratios are compared in the scatter plots of Fig.~\ref{09 scatt} and~\ref{30 scatt}. The $^{13}{\rm CO}(J=1 - 0)$ to $^{13}{\rm CO}(J=2-1)$ (noted $R_{13}$) and ${\rm C^{18}O}(J=1-0)$ to ${\rm C^{18}O}(J=2-1)$ (noted $R_{18}$) ratios are found to be constant on average for a given source, independent of the choice of pixels in the line centre or wings, in the cloud core or envelope. A similar result was found by e.g. Falgarone et al. (1998) for dense cores close to the Sun (d$\lesssim$150 pc). A significant departure from this general behaviour is however observed in the close vicinity of the young stellar objects of DF+30.23--0.20: in the line wings the $R_{13}$ ratio is higher than the average value obtained for the filament (rightmost data points of Fig.~\ref{30 scatt}). The excitation is certainly higher and the $(J=2-1)$ transition is optically thicker than the $(J=1-0)$ one, which results in a broader line and eventually in some self-absorption (Fig.~7).

Table~\ref{ratio} compiles the best-fit results of these ratios. The ratios between integrated intensities are also given for comparison with previous $^{12}$CO studies. These results can be compared to values observed in other clouds of our Galaxy: in quiescent clouds of the solar neighbourhood, Falgarone et al. (1998) found a mean of 0.65$\pm$0.15 for $^{12}$CO and $^{13}$CO. The C$^{18}$O observations reported by Kramer et al. (1999) in IC5146 lead to a mean value of 0.83$\pm$0.22. Several studies have been conducted in the Galactic plane in $^{12}$CO, showing a large-scale gradient of the integrated intensity ratios: 0.75 at $\sim$ 4 kpc from the Galactic centre, to 0.5 at 8 kpc from the Galactic centre (e.g. Sakamoto et al. 1995, who derive a mean value of 0.66$\pm$0.01). A similar study conducted by Oka et al. (1998) in the Taurus and the Orion A Molecular Clouds, including star-forming sites, lead to mean values of 0.53$\pm$0.01 and 0.75$\pm$0.01 respectively.

Our mean values of 0.72$\pm$0.16 ($^{13}$CO) and 0.90$\pm$0.29 (C$^{18}$O) are higher than the values found in these published studies. Even the starless sub-sample shows rather high values: 0.67$\pm$0.12 ($^{13}$CO) and 0.79$\pm$0.22 (C$^{18}$O). Provided relative calibration uncertainties can be excluded, these values may be characteristic of the advanced condensation stage of IRDC's.

The comparison of $^{13}{\rm CO}$ and ${\rm C^{18}O}$ line intensities (called $R_{10}$ and $R_{21}$ in Table~\ref{line param} for $J=1-0$ and $J=2-1$ respectively) illustrate the differential abundances, excitation states and optical depths of the two CO isotopomers. At low intensities (yet above 5$\sigma$), these intensity ratios are consistent with the local interstellar [$^{13}{\rm CO]/[ C^{18}O}$] ratio of about 7.3 (deduced from [$^{16}$O]/[$^{18}$O] = 560 and [$^{12}$C]/[$^{13}$C] = 77, Wilson \& Rood 1994), but they depart from this value as intensities increase. The ratios drop below 2 close to the emission peaks. These ratios are systematically larger in the envelope than in the core areas. The slow apparent saturation of the stronger line should be accounted for by detailed models.

\section{Cloud properties}
\label{properties}
Preliminary results are derived from a simple analysis of the available data: spectroscopic intensities are interpreted with ``LVG'' calculations, which should be seen as the lowest order approximation (1-zone) to radiative transfer calculations.

%%%%%%%%%%%%%%%%%%%%%%%%%%%%%%%%%%%%%%%%%%%%%%%%%%%%%%%%%%%%%%%%%%%%%

\begin{figure*}
 \begin{center}
 \vspace{155mm}
 \includegraphics{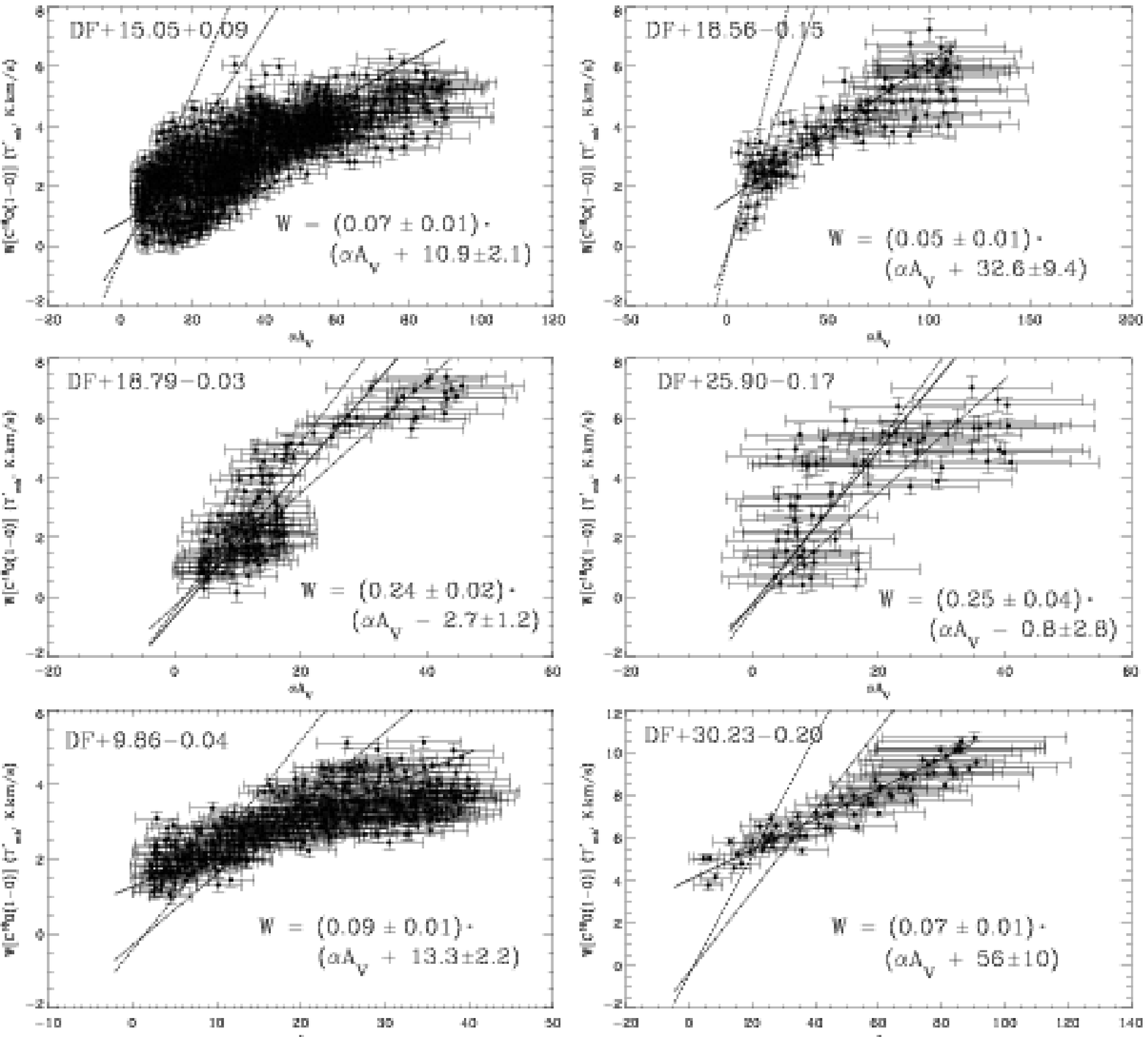}
% \special{psfile=MS10625f12.eps voffset=0 hoffset=10  hscale=73 vscale=73 angle=0}
 \caption{Relation between the C$^{18}$O$(J=1-0)$ integrated intensity and the visual extinction derived from mid-IR measurements in six of the clouds. $\alpha$ depends on the $A_{\rm v}$/$A_{\rm 15 \mu m}$ scaling; $\alpha = 1$ for DL84. IR data were smoothed to the C$^{18}$O beam size of 22\arcsec. The error in the extinction measurements is calculated according to Eq.~\ref{dtau eq} while the uncertainties on W(C$^{18}$O) correspond to the rms noise of the spectra. The solid line represents the least-square fit of the data weighted in both coordinates. The dashed (respectively dashed-dot-dot-dot) lines give the fits obtained for nearby dark clouds by Cernicharo \& Gu\'elin (1987) (resp. Alves et al. 1999) and recalled in Eq.~\ref{Cerni eq} (resp. Eq~\ref{alves eq}).}
 \label{extinction law}
 \end{center}
 \end{figure*}   

%%%%%%%%%%%%%%%%%%%%%%%%%%%%%%%%%%%%%%%%%%%%%%%%%%%%%%%%%%%%%%%%%%%%%

%%%%%%%%%%%%%%%%%%%%%%%%%%%%%%%%%%%%%%%%%%%%%%%%%%%%%%%%

\begin{table*}
\begin{center}
\caption{H$_2$ Column densities averaged over a 20'' beam, in units of $10^{22}$ cm$^{-2}$}
\label{NH2 table}
\begin{scriptsize}
\begin{tabular}{l c c c c c c c} 
\hline \hline 
 & {\bf (1)} & {\bf (2)} & & {\bf (3)} & {\bf (3')} & {\bf (4)} & \\
Source name 		& $N_{\rm H_2}$ (mid-IR) & $N_{\rm H_2}$ (1.2mm) & Assumed  & $N_{\rm H_2}$ (HC$_3$N) & $N_{\rm H_2}$ (C$^{18}$O) & $N_{\rm H_2}$ (C$^{18}$O)& Assumed  \\ 
   		&  &  & $T_{\rm dust}$ (K) & (LVG+size) & (LVG+size) & (Cernicharo et al.)  &$T_{\rm kin}$ (K)\\  \hline

DF+09.86--0.04$^{(A)}$  	& 3.4$\pm$1.5 & 6.1$\pm$1.7 & 10 & -- & 2.9$\pm$1.5 & 1.4$\pm$0.3 & 10\\ 
DF+09.86--0.04$^{(B)}$  	& 3.1$\pm$1.4 & 4.8$\pm$1.7 & 17 & -- & 0.6$\pm$0.1 & 1.1$\pm$0.2 & 17\\ 
DF+15.05+0.09$^{(A)}$ 	& 7.7$\pm$3.4 & 12.6$\pm$1.7 & 8 & -- & 1.2$\pm$0.6 & 1.6$\pm$0.3 & 8\\ 
DF+30.23--0.20 		&  &  &  &  &  &  & \\ 
{\it star vicinity}	& 5.0$\pm$1.8 & 8.3$\pm$2.7 & 25 & 12$\pm$6 & 0.9$\pm$0.1 & 4.2$\pm$0.8 & 25\\ 
{\it filament}		& 8.5$\pm$3.0 & 11.1$\pm$1.7 & 8  & 42$\pm$28& ?$^{(C)}$ & 2.8$\pm$0.5 & 8 \\ 
DF+31.03+0.27 		&  &  &  &  &  &  & \\ 
{\it 1$^{st}$ component$^{(A)}$}  & -- & 11.1$\pm$1.7 & 10 & $\geq$10 & 0.75$\pm$0.15 & 2.0$\pm$0.4 & 10\\ 
{\it 2$^{nd}$ component$^{(A)}$} & -- & 17.7$\pm$1.7 & 10 & 12$\pm$6 & 3.5$\pm$1.5 & 2.0$\pm$0.4 & 10\\ 
DF+51.47--0.00$^{(A)}$ 	& -- & 7.7$\pm$1.7 & 10 & $\geq$14 & 1.2$\pm$0.3 & 1.6$\pm$0.3 & 10\\ \hline

\end{tabular}
\begin{flushleft}
\kern-2mm$^{(A)}$ Taken at {\it core} position\\
$^{(B)}$ Taken in the vicinity of the upper-right star cluster seen in Fig.~\ref{co/iso map}\\
$^{(C)}$ The density could not be constrained by the LVG model
\end{flushleft}
\end{scriptsize}
\end{center} 
\end{table*}

%%%%%%%%%%%%%%%%%%%%%%%%%%%%%%%%%%%%%%%%%%%%%%%%%%%%%%%%%%%%%%%%%%%%%%%%%%%%%%%

\subsection{Temperatures and densities from spectroscopic data}
\label{temp and density}
We used the spectra obtained in the $J=5-4$ and $6-5$ transitions of CH$_3$CCH (Fig.~\ref{hc3n spec}) to derive the kinetic temperatures in a subset of lines of sight. The details of this analysis will be presented in a subsequent paper (Teyssier et al. 2001b). The temperatures inferred are in the range 8-25\,K.
%We obtained spectra of the CH$_3$CCH symmetric top molecule in the $J=5-4$ and $6-5$ transitions (see Fig.~\ref{hc3n spec}). Using the charts presented by Askne et al. (1984) and Pratap et al. (1997) we compared our results to the $(K=1)/(K=0)$ and $(K=2)/(K=1)$ ratios they derived from LVG simulations. Assuming that these molecules trace the densest parts of our clouds, we can infer temperatures in the 8-30 K range. 
The highest temperatures are representative of clouds with embedded stars (e.g. DF+30.23--0.20). These are also the only areas where significant emission is detected in the $K=2$ rotation level.

We use these temperatures to derive densities and molecular column densities from LVG simulations of HC$_3$N, $^{13}$CO and C$^{18}$O. The HC$_3$N transitions are relatively close in energy so that the error bars are somewhat large. The densities derived are larger than $10^{5}$ cm$^{-3}$ in the densest parts ({\it cores}). Observations of lower frequency transitions are included for more accurate derivations in our subsequent paper (Teyssier et al. 2001b). The corresponding masses vary between $2\times10^{2}$ and $2\times 10^{4}$ $\rm M_\odot$. The temperatures and densities of our sample clouds are similar to the parameters reported by Carey et al. (1998) for the MSX IRDC's.

\subsection{Column density comparison}
\label{col dens}
For several of the sources we can compare independent estimates of total column densities: from mid-IR opacities, from dust 1\,mm emission, from densities and size estimates, or from the column densities of tracers.

The first method consists in relating the mid-IR opacities to visible extinctions, hence total gas column density. The dust emissivity model of Draine \& Lee (1984, hereafter DL84) implies:
\begin{eqnarray}
 A_{\rm v} \simeq 65 \times \tau_{7\mu{\rm m}}
\label{d&l eq}
\end{eqnarray}

%In Paper I a systematic analysis of the ISOGAL data leads to opacities at 7 and 15 $\mu$m. The method is consistent with a $\tau_{\rm 7 \mu m}/\tau_{\rm 15 \mu m}$ ratio of $0.65 \pm 0.05$, in good agreement with results extrapolated from dust models by Draine \& Lee (1984, hereafter DL).  For this we used an average of the maximum opacities obtained for $\tau_{\rm 7 \mu m}/\tau_{\rm 15 \mu m}$ ratios of either 0.6 or 0.7. Assuming that $A_{\lambda} \simeq \tau_{\lambda}$ at these wavelengths, one estimates the visible extinction from the relation:

\noindent The total column density along the line of sight can then is estimated using the ($N_{\rm H+H_2}$,${\rm A_V}$) relation of Bohlin et al. (1978). 

The second method is based on continuum measurements of the optically thin dust at 1.2\,mm and uses the relation described in Motte et al. (1998):
\begin{eqnarray}
N_{\rm H_2} = \frac{S_{\rm 1.2mm}}{\Omega_{\rm beam}\,\mu\,{\rm m_H}\,\kappa_{1.2}\,B_{1.2}({\rm T}_{dust})}
%N_{\rm H_2} \simeq 4\times10^{20}\times S_{\rm mm} \times 
%\Big(\frac{J({\rm T_{dust},\nu})}{20\,{\rm K}} \Big)^{-1}~~{\rm cm^{-2}}
\label{motte eq}
\end{eqnarray}

\noindent where $S_{\rm mm}$ is the 1.2\,mm flux density in mJy per 11\arcsec~beam, $\mu$ the mean molecular weight, m$_{\rm H}$ the mass of atomic hydrogen, $B_{1.2}(T)$ the Planck function at wavelength 1.2\,mm and temperature T, and $\kappa_{1.2}$ the dust mass opacity at 1.2\,mm. We adopted a value $\kappa_{1.2}= 0.003$ cm$^2$\,g$^{-1}$, consistent with the DL84 dust model and the ($N_{\rm H+H_2}$,${\rm A_V}$) relation used in method {\bf (1)}. Such a value is believed to be suited to pre-stellar dense clumps and cores with a typical uncertainty of a factor of $\sim$ 2 (see Motte et al. 1998 and ref. therein for a complete discussion). The dust temperature is assumed to be equal to the gas temperature as estimated in section~\ref{temp and density}.

In the third method we assume that the H$_2$ density derived from beam-averaged spectra is representative of the density within a sphere of projected size in the sky equal to the beam diameter. This assumption seems viable, in view of the apparent size of the core regions, but might be very uncertain, if the gas distribution is lacunar. Estimates are given both from HC$_3$N (method {\bf (3)}) and from C$^{18}$O (method {\bf (3')}).

Finally (method {\bf (4)}), we used the ($N_{\rm H_2}$,$W$(C$^{18}$O)) relation derived by Cernicharo \& Gu\'elin (1987, hereafter CG87) in local dark clouds. Table~\ref{NH2 table} gathers our results.

The column densities range from a few $10^{22}$ to at most a few $10^{23}$ cm$^{-2}$.
The ratio of the estimates from methods {\bf (1)} and {\bf (2)} is within $\pm$ 15\% of 0.65 for all 4 cases where both estimates are available. A higher value of the emissivity at 1.2\,mm (1.6 times the DL84 value) is required for matching the 2 derivations. This agrees with observations of cold condensations in the solar neighbourhood (e.g. Dupac et al. 2001). A proper account of the dust temperature distribution is in all cases requested for deriving more accurate conclusions.

Estimates from method {\bf (4)} are 2 to 9 times smaller than values traced by the dust, the better agreement corresponding to the embedded star cluster of DF+30.23--0.20. Since most of the reported points have high $A_{\rm v}$, this trend suggests possible molecular depletion onto very cold grains(see next section).

Since we considered the same core size for both cases {\bf (3)} and {\bf (3')}, the discrepancy between these two estimates is not surprising: it illustrates how the density derivations are limited by the different critical densities (e.g. Evans 1980). 
On the other hand, the column densities derived from HC$_3$N ({\bf (3)}) are generally higher than those inferred from the dust ({\bf (1),(2)}). This may indicate that the actual volume filling factor is smaller than 1, the value assumed for this estimate. As in the picture proposed by Lada, Evans \& Falgarone (1997) in local dense cores, this result suggests that the densest regions of the clouds are further fragmented.

\subsection{Extinction}
\label{extinction}

Using Eq.~\ref{d&l eq}, we studied the pixel-to-pixel correlation between visual extinction and the C$^{18}$O$(J=1-0)$ integrated intensity for selected clouds. The error estimate on the mid-IR opacities is described in appendix~B.
 Fig.~\ref{extinction law} shows the resulting scatter diagrams. The scaling of extinction data to $A_{\rm v}$ depends on the assumed dust emissivity model. We then denote the visual extinction scale $\alpha A_{\rm v}$, where $\alpha = 1$ for the DL84 model. The result of bivariate fits of our data are displayed on Fig.~\ref{extinction law}. The errors resulting from the linear regression show that this best fit line is relatively well constrained, regardless of its 
adequacy, or not, to properly represent the data over the whole $A_{\rm v}$ range, of the intrinsic data dispersion and of the absolute calibration uncertainty.

We now compare our results to earlier studies of local dark clouds, based on star counts. In HCL2, CG87 inferred for $A_{\rm v}$ ranging from 1.5 to 6:
\begin{eqnarray}
W({\rm C^{18}O}) = 0.28\pm0.05 \cdot (A_{\rm v} - 1.5\pm0.3) \; {\rm K~km~s^{-1}}
\label{Cerni eq}
\end{eqnarray}
For $A_{\rm v} \leq 10$, Alves et al. (1999) found in L977:
\begin{eqnarray}
W({\rm C^{18}O}) = 0.18\pm0.01 \cdot (A_{\rm v} - 1.67\pm0.26) \; {\rm K~km~s^{-1}}
\label{alves eq}
\end{eqnarray}
These relations are overplotted on the graphs of Fig.~\ref{extinction law}. An important difference with the earlier studies is that the extinction ranges probed barely overlap. 
%[\sout{A second difference lies in}]
The different location and nature of the sampled clouds observed with different spatial resolutions might lead to large discrepancies. The different data sets nevertheless seem compatible. Our scatter diagrams exhibit a similar break as reported by Alves et al. (1999, see also Frerking et al. 1982, Lada et al. 1994) above $A_{\rm v}\sim$ 10-15 mag. In our case, the departure from the linear fit occurs at higher $A_{\rm v}$ in the range 15-25 magnitudes, which is about the sensitivity limit of the previous studies. It is interesting to note that the relations given in Eqs.~\ref{Cerni eq} and~\ref{alves eq} are not inconsistent with our few sample points at $A_{\rm v}$ below 15-20 magnitudes (e.g. DF+18.56--0.15, DF+25.90--0.17).

As for previous properties, our clouds are not all identical and their behaviour in the ($W({\rm C^{18}O}),A_{\rm v}$) plane may characterise different 
%[\sout{situations}]
conditions:

1. {\it DF+15.05+0.09} and {\it DF+18.56--0.15}: these clouds are identified as the most opaque objects of the sample and thus suffer most from the mismatch in the extinction ranges. Nevertheless the low extinction relations (Eqs.~\ref{Cerni eq} and~\ref{alves eq}) smoothly connect to our data points and are consistent with a break at $A_{\rm v}$ around 15, similar to Alves et al. (1999) findings. Their conjecture that such a break is due to molecular depletion onto grains is validated by the Kramer et al.'s (1999) study based on rare CO isotopomers. The similar break found here reinforces our assumption of significant molecular depletion inside the IR dark clouds.

2. {\it DF+18.79--0.03} and {\it DF+25.90--0.17}: these clouds present much more points in the extinction range of Eqs.~\ref{Cerni eq} and~\ref{alves eq}. Although the samples are relatively scattered,
%[\sout{(respectively 155 and 71 positions)}]
we find that the data are nicely fitted by the relation of CG and Alves et al. (1999) within 10-30\%, which is still consistent given the uncertainty implied by the factor $\alpha$ introduced above. The break position is, again, consistent with previous results and the negative intercept in agreement with Alves et al. (1999) claim that shielding is probably not efficient enough to avoid molecular photodissociation from interstellar radiation at low visual extinctions.

3. {\it DF+9.86--0.04} and {\it DF+30.23--0.20}: these clouds exhibit a clear lack of points of weak C$^{18}$O emission at low visual extinctions. As mentioned in previous sections, these objects are associated with several young star clusters whose radiation probably heatens the surrounding core. In DF+30.23--0.20, several OH and CH$_3$OH masers (Caswell et al. 1995) have been identified at the edge of the filament (see Fig.~\ref{bolo maps}), probably tracing an even more evolved state of star-formation.

We conclude that our study extends earlier studies based on star counts in a consistent way. In particular it confirms that C$^{18}$O can be used as a rough column density, and thus mass, tracer in dense dark clouds, provided corrections are made to account for progressive saturation.

\section{Conclusions}
\label{conclusion}
In this paper we presented millimetre follow-up observations of Galactic infrared dark clouds discovered in the galactic ISO survey at 7 and 15 $\mu$m. All objects are detected in $^{13}$CO and C$^{18}$O and present remarkable spatial correlation with the mid-IR absorption data. We found that these clouds are not isolated but appear as the most condensed parts of quiescent GMC's 2 to 8 kpc away from us, some of them being associated with embedded stars or clusters observed by ISO. The clouds present a variety of shapes and line profiles indicative of supersonic flows. Velocity maps exhibit structures down to the parsec size connected to larger clouds along the line of sight.

Special attention was paid to the intensity calibration of our data and a simple approximation for extended sources seen in a main+error beam pattern was developed. Although not completely correct, it is believed to provide a better estimate of the main error beam integrated intensity than the widely used main-beam brightness temperature.
 
In spite of a certain diversity, some common behaviours are observed in our cloud sample. We found that the $(J=2-1)$ to $(J=1-0)$ ratios of $^{13}$CO and C$^{18}$O were remarkably uniform within each object and that a significant amount of them match the values between 0.6 and 0.8 observed in local dark clouds. Higher ratios are observed in the other objects, indicative of a broader range of conditions.

A preliminary analysis of data obtained with temperature and density probes such as HC$_3$N and CH$_3$CCH shows that the clouds are mostly cold gas and dust condensations of temperatures below 10 K and densities in excess of $10^{5}$ cm$^{-3}$. In those conditions, depletion onto grains is expected to strongly affect the molecular emission, but opacity effects can so far not be completely ruled out. Column densities close to $10^{23}$ cm$^{-2}$ are inferred from the dust absorption and emission. These clouds are in an advanced condensation stage, and some of them already contain young stars. Similar massive dark clouds are reported by Egan et al. (1998) and Carey et al. (1998) in their MSX survey of the Galactic plane: the parameters they derive are in good agreement with ours.

The darkest condensations ($N_{\rm H_2} \ge 10^{23}$ cm$^{-2}$) represent $\sim$\,1\% of the inner Galaxy surveyed with ISO. Assuming that $\lesssim$~80\% of the population escaped detection (distance bias, confusion,..), a wild extrapolation leads to at most 10\% of the dust in the inner Galaxy trapped in these cold dense cores. This is a factor $\sim$~50 too low to account for the hypothetical cold dust component advocated by Reach et al. (1995).

Comparison of dust mid-IR absorption and 1.2\,mm emission points towards emissivities at 1.2\,mm higher (factor $\sim$ 1.6) than Draine \& Lee (1984). On the other hand spectroscopic tracers indicate that the denser parts of the cold condensations are not volume filling, as earlier found in other situations (e.g. Lada et al. 1997).

We studied the relation between the integrated C$^{18}$O emission and the visual extinction and compared it to previous results obtained for local clouds. Our study samples higher extinctions than these former studies but we find that the correlations consistently connect in the $A_{\rm v}$ = 15-25 mag range and indicate that the C$^{18}$O emission still is a relatively good tracer of the cloud mass at visual extinction above 15-20 magnitudes, provided corrections are made to account for radiative transfer saturation and/or depletion onto grains.

%One of the major issue to address is whether or not these clouds could be preferential sites for star formation. Young star clusters and masers are indeed associated to some of the objects. The question of the gravitational support of such massive objects should be investigated and the contribution of actors such as macro-turbulence or magnetic field needs to be assessed.

\begin{appendix}

\section{Observation and reduction techniques}
\label{append 1}

\subsection{Brightness temperature scaling}
\label{append 1.1}
When observing extended non-uniform emission, a significant fraction of the signal is peaked by diffraction pattern components larger than the main beam. As a consequence, the widely used main beam brightness temperature scale (${ T_{\rm mb}}$, also defined as ${ T^*_{\rm R}}$ by Kutner \& Ulich, 1981) over-estimates the actual physical line intensity derived from the observations. Several methods have been proposed to correct from these effects (see Bensch et al. 2001a for a recent review) but most of them require additional observations to be conducted on larger scale and generally smaller telescopes.

In this work, we have used an approximation which consists of assimilating the source to a uniform disk-like emitter of brightness temperature ${ T_{\rm b}}$. If the complete beam of the instrument is known, it is possible to calculate the ratio between ${ T_{\rm mb}}$ and the real ${ T_{\rm b}}$ as a function of the source size. The 30-m beam is modelled according to Greve et al. (1998) as the superimposition of a main beam and three extended error beams:
\begin{eqnarray}
P = \sum_{\rm i=0,3}~P_{\rm i} =
\sum_{\rm i=0,3}~\frac{w_{\rm i}}{\Omega_{\rm i}}~{\rm exp[-ln2~(2\theta / \theta_i)^2]} ~d\theta
\label{eq beam pattern 1}
\end{eqnarray}
where $P_0$ is the main beam beam pattern and
\begin{eqnarray}
w_{\rm i} = \int\!\!\!\! \int _{2\pi} P_{\rm i}~d\Omega
\label{eq omega i}
\end{eqnarray} 
is the power fraction contained in the beam component $P_{\rm i}$, and $\Omega_{\rm i}$ its corresponding solid angle.
Using the definition by Kutner \& Ulich (1981), the main beam brightness temperature can be written:
\begin{eqnarray}
T_{\rm mb} = \frac{\int \!\!\int _S~T_{\rm b}\star P~d\Omega}{\int \!\!\int _{\rm mb}~P~d\Omega}
\label{eq Tmb}
\end{eqnarray}
For Gaussian beams and a disk-like uniform source of diameter $\theta_{\rm s}$ and temperature $T_{\rm b}$, Eq.~\ref{eq Tmb} finally is written:
\begin{eqnarray}
T_{\rm mb} = \frac{T_{\rm b}}{\beta_{\rm 0}}~
		\left[1-\sum_{\rm i=0,3}\beta_{\rm i}~
		{\rm exp[-ln2~(\theta_s/\theta_{\rm i}})^2]\right]
\label{eq Tmb 2}
\end{eqnarray}
with
\begin{eqnarray}
\beta_{\rm i} \times \int\!\!\!\! \int _{2\pi} P~d\Omega = w_{\rm i}~~{\rm (Panis~1995)}
\end{eqnarray} 
Similarly, one has:
\begin{eqnarray}
T^*_{\rm A} = \beta_{\rm 0}~T_{\rm mb} {\rm ,~where~}\beta_{\rm 0} = \frac{B_{\rm eff}}{F_{\rm eff}}
\label{eq Ta*}
\end{eqnarray}
is the ratio of the main beam to forward efficiency (e.g. Kramer et al. 1997). Table~\ref{disklike result} lists the ratios for various disk diameters. As expected, effects at 3 mm remain negligible if the source is smaller than the first error beam diameter.

To estimate the accuracy of this crude approximation, we compared it to a more accurate correction used by Falgarone et al. (1998) and Bensch et al. (2001b) on their data. We found an agreement within 10\% for the brightness temperatures at 1.3 mm, where the error beam contamination is rather significant, and within 7\% at 3mm, where the error beam peak-up is much smaller.

For our data, the corrections to $T_{\rm mb}$ amount to 5-7 \% at 3 mm and to values in the range 10-70 \% at 1.3 mm. Emission at a larger distance from the observed position, which is not considered here, might increase the 3 mm correction a little. For most sources the adopted correction at 1.3 mm corresponds to emission extending beyond the first error beam ($\sim$ 2\arcmin~diameter). One of the greatest limitations of the method shows up for elongated sources but still the geometrical mean should represent {\it on average} the fraction of the source surface picked up beyond the main beam. Despite this sometimes severe limitation, we conclude that our approximate correction reasonably accounts for error beam pick-up and that the residual error on $T_{\rm b}$ is significantly smaller than the error resulting from using either $T_{\rm mb}$ or $T^*_{\rm A}$.

%%%%%%%%%%%%%%%%%%%%%%%%%%%%%%%%%%%%%%%%%%%%%%%%%%%%%%%%%%%%

\begin{table}
\begin{center}
\caption{Error on brightness temperature for a disk-like uniform source of diameter $\theta_{\rm s}$}
\label{disklike result}
\begin{small}
\begin{tabular}{c c c c c} 
\hline \hline \\
$\theta_{\rm s}$ & \multicolumn{2}{c}{3 mm} & \multicolumn{2}{c}{1.3 mm} \\
(\arcsec)     &  $T_{\rm mb}/T_{\rm b}$ & $T^*_{\rm A}/T_{\rm b}$  
& $T_{\rm mb}/T_{\rm b}$ & $T^*_{\rm A}/T_{\rm b}$\\  \hline
26.2	&	0.67	&	0.47	&{\bf	1.00}	&	0.42	\\
30  	&	0.77	&	0.54	&	1.02	&	0.43	\\
46  	&{\bf	1.00}	&	0.70	&	1.06	&	0.45	\\
60  	&	1.04	&	0.73	&	1.10	&	0.46	\\
90  	&	1.05	&	0.74	&	1.20	&	0.51	\\
120 	&	1.05	&	0.74	&	1.32	&	0.56	\\
150  	&	1.06	&	0.74	&	1.44	&	0.60	\\
180 	&	1.07	&	0.75	&	1.54	&	0.65	\\
240 	&	1.08	&	0.76	&	1.68	&	0.71	\\
600 	&	1.14	&	0.80	&	1.90	&	0.80	\\
6000	&	1.43	&{\bf	1.00}	&	2.38 	&{\bf	1.00}	\\ \hline 
\end{tabular}
\end{small}
\end{center} 
\end{table} 

%%%%%%%%%%%%%%%%%%%%%%%%%%%%%%%%%%%%%%%%%%%%%%%%%%%%%

\subsection{Spurious effects in the spectral baseline}
\label{append 1.2}
Our spectral baselines are affected by standing waves (resulting from multiple reflexions along the optical paths) and by discontinuities in the autocorrelator (AC) baseline (gain mismatch between sub-bands, also known as {\it platforming}). 
The platforming can be efficiently corrected if a wider band backend (here a filter bank) is connected in parallel. This permits us to provide an accurate 0-level to each sub-band, independent of baseline ripples that are seen equivalently in both spectrometers. A sub-band readjustment based on 0-order baselines separately estimated in each sub-band of the AC alone would indeed alter the wideband standing wave pattern and would thus bias any further correction of this additional effect.

\section{Error calculation for mid-IR opacities}
\label{append 2}
The error estimate on the opacities calculated in Paper I is based on the following formalism for the mid-IR emission:
\begin{eqnarray}
I_{\rm tot} = I_{\rm bg}~ {\rm e}^{-\tau} + I_{\rm fg}
\label{pat formalism eq}
\end{eqnarray}
where $I_{\rm bg}$ and $I_{\rm fg}$ are respectively the background and foreground emission w.r.t. the cloud of opacity $\tau$. We assume that the main uncertainty on $\tau$ comes from the intrinsic fluctuations of $I_{\rm bg}$ and $I_{\rm fg}$, of total rms amplitude $\epsilon$, which we estimate in unobscured neighbouring regions. Using the parameter $\beta$ introduced in Paper I:
\begin{eqnarray}
\beta = \frac{I_{\rm bg}}{I_{\rm tot}}
\label{beta eq}
\end{eqnarray}
the fluctuations of $I_{\rm bg}$ and $I_{\rm fg}$ are written:
\begin{eqnarray}
\epsilon_{\rm bg} = \sqrt{\beta}~\epsilon,~~{\rm and}~~
\epsilon_{\rm fg} = \sqrt{1-\beta~{\rm e}^{-\tau}}~\epsilon
\label{eps fg and bg eq}
\end{eqnarray}
Assuming that these contributions are uncorrelated, this leads to:
\begin{eqnarray}
\delta \tau^2 & \simeq & \Big(\frac{\epsilon_{\rm bg}}{I_{\rm bg}}\Big)^2 +
		    \Big(\frac{\epsilon_{\rm fg}~{\rm e}^{\tau}}{I_{\rm bg}}\Big)^2 \\
	      & \simeq & \Big(\frac{\epsilon}{I_{\rm tot}}\Big)^2~\Big[
		\frac {1}{\beta}+\frac{1-\beta~{\rm e}^{-\tau}}{\beta^2}~
		({\rm e}^{\tau})^2 \Big]
\label{dtau eq}
\end{eqnarray}

This approach tends to favour the lowest extinction points of our samples but takes into account the fact that errors on nearby objects ($\beta \lesssim 1$) are smaller.

\end{appendix}

\begin{acknowledgements} We thank the anonymous referee for constructive comments and suggestions that improved the paper. We also would like to thank Christoph Nieten for providing us with a modified version of the FLYPLAIT
program for data reduced with the GILDAS software. D. Teyssier is very grateful to the IRAM 30-m staff for his help and assistance during his period there. We also acknowledge P. Schilke for providing his LVG code. We would like to thank J.F. Panis for the very useful discussions on temperature scaling and for making the IRAM key-project raw database available to us. F. Zagury of Nagoya University made the observations at the 4-m Nanten telescope. 
\end{acknowledgements}

{}

\end{document}